\documentstyle[12pt,epsf]{article}
\begin{document}

\title{Generalized Jacobi Identities, Curvature Relation, 
Schouten's Identity, a Phase Rule and Derivation of  O$(q^4)$ 
Effective Lagrangian in the Presence of External Fields
Directly Within Heavy Baryon Chiral Perturbation Theory} 
\author{A. Misra\thanks{e-mail: aalok@iopb.res.in},\\
Institute of physics, Sachivalaya Marg,\\
Bhubaneswar 751 005, Orissa, India}
\maketitle
\vskip 0.5 true in

\begin{abstract}
This is a progress report on the extension of the analysis of 
\cite{O4ext0} to  constructing a complete list of O$(q^4)$ terms
in the presence of external fields, and including isospin-violation
directly within Heavy Baryon Chiral Perturbation Theory (HBChPT) 
{\it without having to first construct the relativistic BChPT Lagrangian
and then carry out the ${1\over\rm m}$-reduction}.
In addition to a phase rule to implement all symmetries including
charge conjugation invariance directly at the nonrelativistic level,
generalized  Jacobi identities, curvature relation
(that relates the commutator of two covariant derivatives
to a linear combination of the traceless and
isosinglet field strengths and the commutator of two
axial vector building blocks), Schouten's identity
and the relationship between the antisymmetrized
covariant derivative - axial-(building block) vector 
commutator and another traceless field strength, are used to ensure 
linear independence of the terms and their low energy coupling
constants.  We first construct O($q^4$) terms
for off-shell nucleons,  and then perform the on-shell reduction,
again within HBChPT. 
\end{abstract}



\clearpage 

\section{Introduction}

Heavy Baryon Chiral Perturbation Theory (HBChPT) is a nonrelativistic
(with respect to the ``heavy" baryons) effective field theory
used for studying meson-baryon interactions at low energies, typically
below the mass of the first non-Goldstone resonance
(See \cite{jm,bkm,bkm1}). The
degrees of  freedom of SU(2) ($\equiv$ isospin)  HBChPT 
(which will be considered in this paper) are the (derivatives of) pion-triplet, 
the nucleon fields and the external fields.

Recently, a method was developed to generate HBChPT Lagrangian 
(${\cal L}_{\rm HBChPT}$) for off-shell nucleons {\it directly within HBChPT},
which as stated in \cite{1n},
can prove useful when applying HBChPT  to nuclear processes in which
the nucleons  are bound, and hence off-shell. This method has the
advantage of not having to bother to start with the relativistic BChPT
Lagrangian and then carry out the nonrelativistic reduction. It
is thus shorter than the standard approach to HBChPT
as given in \cite{em}, showed explicitly up to O$(q^3)$ in \cite{1n}.
In the context of off-shell nucleons, the upshot of
the method developed is a phase rule (See (\ref{eq:phoffsh}),\cite {1n})
to implement charge conjugation  invariance (along
with Lorentz invariance, parity and hermiticity) 
directly within HBChPT.  The phase rule, along with additional
reductions from a variety of algebraic identities, was used to construct,
directly within HBChPT, a complete list  of off-shell O($q^3$) terms 
(in the isospin-conserving approximation and in the abscence of external
fields). We also showed
that the  on-shell limit of the list of terms obtained
matches the corresponding list in \cite{em} (in which the HBChPT Lagrangian
up to O$(q^3$) was constructed starting
from the relativistic BChPT Lagrangian). 

For a complete and precise calculation in the single-nucleon sector to one loop,
e.g., 1-loop corrections to pion production
off a single (on-shell) nucleon, because of convergence problems 
(assocated with the amplitude ``$D_2$" for
pion production off a single nucleon), one needs to 
go up to O$(q^4)$ (See \cite{bkm1,bkm2,mms}).
An overcomplete list of the {\it divergent} O($q^4$) 
$\pi$-nucleon interaction  terms in the
presence of external fields was constructed in \cite{mms}, 
but again starting from the relativistic
BChPT Lagrangian. In \cite{O4ext0}, we constructed a complete list of 
O$(q^4)$ terms {\it working entirely within  the framework  of
HBChPT} in the absence of external fields  
and in the isospin-conserving approximation. In this paper,
we extend the list to include  external  fields
and do not assume isospin symmetry.

Section 2 has the basics and sets up the notations. In Section 3, 
reductions obtained in addition  to (\ref{eq:phoffsh}) 
due to algebraic
identities such as generalized Jacobi identities, Schouten's identitiy and
curvature relation, etc. are discussed.
In Section 4, the complete lists
of O$(q^4)$ terms is given. 
In Section 5 we discuss
the derivation of the external field-dependent 
on-shell O$(q^4$) Lagrangian, again within HBChPT
using the techniques of \cite{1n}. Section 6 has the conclusion 
and discussion on comparison of the results of this paper with those
of a recent paper by Fettes et al \cite{Fettes}, in which
the O$(q^4)$ list in the presence of external fields is derived, but
starting from the relativistic O$(q^4$) BChPT Lagrangian.

\section{The Basics}

The HBChPT Lagrangian is written in terms of the ``upper component"
H (and its hermitian adjoint ${\bar{\rm H}}$), 
exponentially parameterized matrix-valued meson fields $U,\ u\equiv\sqrt{U}$, 
baryon (``$v_\mu,\rm S_\nu$") and
pion-field-dependent (``${\rm D}_\mu, u_\nu,\chi_\pm$,
$F^{\pm}_{\mu\nu},v^{(s)}_{\mu\nu}$") building
blocks defined below:
\begin{equation}  
\label{eq:Hdef}
{\rm H}\equiv e^{i{\rm m}v\cdot x}{1\over2}(1+\rlap/v)\psi,
\end{equation}
where $\psi\equiv$ Dirac spinor
and $m\equiv$ the nucleon mass; 
\begin{eqnarray} 
\label{eq:Sdef}
& & 
v_\mu\equiv{\rm nucleon\ veclocity},\nonumber\\
&  & {\rm S}_\nu\equiv{i\over2}\gamma^5\sigma_{\nu\rho}v^\rho
\equiv{\rm Pauli-Lubanski\ spin\ operator};
\end{eqnarray}
\begin{equation}
\label{eq:Udef}
U={\rm exp}\biggl(i{\phi\over F_\pi}\biggr),\ {\rm where}\
\phi\equiv\vec\pi\cdot\vec\tau,
\end{equation}
where $\vec\tau\in$ nucleon isospin generators;
${\rm D}_\mu=\partial_\mu+\Gamma_\mu-iv^{(s)}_\mu$ where 
$\Gamma_\mu\equiv {1\over 2}\biggl((u^\dagger(\partial_\mu-ir_\mu)u
+u(\partial_\mu-il_\mu)u^\dagger\biggr)$
($v^{(s)}_\mu$ is the isosinglet vector field needed to generate the
electromagnetic current (See \cite{em})); 
$u_{\mu}\equiv i\biggl(u^{\dagger}({\partial}_{\mu}-ir_\mu)u - 
u({\partial}_{\mu}-il_\mu)u^{\dagger}\biggr)$; 
$\chi_\pm\equiv u^\dagger\chi u^\dagger\pm
u\chi^\dagger u$, where $\chi\equiv 2B(s+ip)$,$s-{\cal M}(\equiv$ quark
mass matrix) and $p$ 
being the external scalar and pseudo-scalar fields;
$F^\pm_{\mu\nu}\equiv u^\dagger F^R_{\mu\nu}u\pm uF^L_{\mu\nu}u^\dagger$ where
$F^R_{\mu\nu}\equiv \partial_{[\mu}r_{\nu]}-i.[r_\mu,u_\nu]$,
and $F^L_{\mu\nu}\equiv\partial_{[\mu}l_{\nu]}-i[l_\mu,l_\nu]$ and 
$v^{(s)}_{\mu\nu}\equiv\partial_{[\mu}v^{(s)}_{\nu]}$ in 
which $r_\mu\equiv V_\mu+A_\mu,\ l_\mu\equiv V_\mu-A_\mu$,
where $V_\mu, A_\mu$ are external vector and axial-vector fields.

Terms of the ${\cal L}_{\rm (H)BChPT}$ constructed from products of 
building blocks will automatically be chiral invariant. 
Symbolically, a term in ${\cal L}_{\rm HBChPT}$ 
can be written as just a product of the building blocks to
various powers (omitting $\rm H,{\bar{\rm H}}$ as will be done in the
rest of the paper except for Section 5):
\begin{equation} 
\label{eq:prodbb}
{\rm D}_\alpha^m 
u_\beta^n\chi_+^p
\chi_-^q v_\sigma^l(v^{(s)}_{\sigma\omega})^k
(F^+_{\rho\lambda})^t(F^-_{\mu\nu})^u
\rm S_\kappa^r\equiv(m,n,p,q,t,u,k)
\equiv{\rm O}(q^{m+n+2p+2q+2k+2t+2u}).
\end{equation}

A systematic path integral derivation for ${\cal L}_{\rm HBChPT}$ based
on a paper by Mannel et al \cite {mnnl}, starting from
${\cal L}_{\rm BChPT}$ was first given by Bernard et al \cite {bkm2}.
As shown by them, after integrating
out h from the generating
functional, one arrives at ${\cal L}_{\rm HBChPT}$ :
\begin{equation}
\label{eq:lag}
{\cal L}_{\rm HBChPT} = {\bar{\rm H}}\biggl( {\cal A} +
\gamma^0 {\cal B}^\dagger\gamma^0 {\cal C}^{-1}{\cal B}
\biggr)\rm H,
\end{equation}
an expression in the upper components only i.e. for non-relativistic
nucleons. So the terms of ${\cal L}_{\rm HBChPT}$ in this paper are 
given as operators on the H-spinors. For off-shell nucleons,
$\gamma^0 {\cal B}^\dagger\gamma^0 {\cal C}^{-1}{\cal B}\in {\cal A}$, 
and hence, listing
${\cal A}$-type terms will suffice.

The phase rule derived in \cite {1n} can be modified to include external fields. 
After doing so, one gets:
HBChPT terms (that are Lorentz scalar - isoscalars of even parity) made 
hermitian using a prescription for constructing hermitian (anti-)commutators
discussed in \cite {1n}, consisting of $q\ \chi_-$'s,
$P[\ ,\ ]$'s, $j$ (which can take only the values 0 or 1) 
$\epsilon^{\mu\nu\rho\lambda}$'s, $k v^{(s)}_{\mu\nu}$'s, $t F^+_{\rho\lambda}$'s
and $u F^-_{\mu\nu}$'s, for which 
the following phase rule is satisfied, are the only terms allowed:
\begin{equation} 
\label{eq:phoffsh}
(-1)^{q+P+j+k+t+u}=1.
\end{equation}
In \cite {1n} for $k=t=u=0$, (\ref{eq:phoffsh}) 
was used to generate complete lists up to O$(q^3)$ in
the absence of external vector and axial-vector fields. In
this paper, the same phase rule is used to construct complete lists of 
O$(q^4)$ including external fields. 

Let $A,B,C,D$ be operators chosen from the pion-field dependent
building blocks of (\ref{eq:prodbb}).  In what follows, and
especially in Section 3, use will be made of a notation of \cite{krause}:
$(A,B)\equiv [A,B]$ or $[A,B]_+$. One can
then show that apart from the  
(0,0,0,2,0,0,0)-,
(0,0,2,0,0,0,0)-, (0,0,0,0,2,0,0)-, (0,0,0,0,0,0,2)-
and (0,0,0,0,1,0,1)-type
terms (using the notation of  (\ref{eq:prodbb})),
the following is the  
complete list of  O$(q^4,\phi^{2n})$ terms (using (\ref{eq:phoffsh})):
\begin{eqnarray}  
\label{eq:list}
& (i) & (A,(B,(C,D)))\equiv
(a)[A,[B,[C,D]_+]];\ (b)[A,[B,[C,D]]_+];\nonumber\\
& & (c)[A,[B,[C,D]]]_+;\ (d)[A,[B,[C,D]_+]_+]_+; \nonumber\\ 
& (ii) & ((A,B),(C,D))\equiv  
(a)[[A,B],[C,D]_+];\ (b)[[A,B]_+,[C,D]];\ \nonumber\\
& & (c)[[A,B],[C,D]]_+;\ (d)[[A,B]_+,[C,D]_+]_+;\ \nonumber\\
& (iii) & i(A,(B,(C,D)))\equiv
(a)i[A,[B,[C,D]_+]_+];\ (b)i[A,[B,[C,D]_+]]_+;\ \nonumber\\
& & (c)i[A,[B,[C,D]]_+]_+;\ (d)i[A,[B,[C,D]]];\ \nonumber\\
& (iv) & i((A,B),(C,D))\equiv  
(a)i[[A,B],[C,D]];\ (b)i[[A,B]_+,[C,D]_+];\ \nonumber\\
& & (c)i[[A,B],[C,D]_+]_+;\ (d)i[[A,B]_+,[C,D]]_+;\nonumber\\
& (v) & i(A,(B,C))\equiv (a)i[A,[B,C]]_+;\ (b)i[A,[B,C]_+];\ (c)A\leftrightarrow B;\nonumber\\
& & (d)i[[A,B],C]_+;\ (e) i[[A,B]_+,C];\ \nonumber\\
& (vi) &  (A,(B,C))\equiv (a)[A,[B,C]];\ (b)[A,[B,C]_+]_+;\ (c)A\leftrightarrow B\nonumber\\
& & (d)[[A,B],C];\ (e)[[A,B]_+,C]_+,
\end{eqnarray}
where it is understood that of all the possible
terms implied by $(i)(A,(B,(C,D)))$, $(i)((A,B),(C,D))$ and
$(i)(A,(B,C))$, only those that are allowed by (\ref{eq:phoffsh}) are
to be included. For (0,0,0,2,0,0,0)-,
(0,0,2,0,0,0,0)-, (0,0,0,0,2,0,0)-, (0,0,0,0,0,0,2)-
and (0,0,0,0,1,0,1)-type terms, one needs to\\
include
\footnote{One need not consider (0,0,0,0,0,2,0). See (10) and
the discussion thereafter.}
\begin{eqnarray}
\label{eq:chi+-}
& & 
(A,A)
\equiv(\chi_+,\chi_+);\ (\chi_-,\chi_-);\ (F^+_{\mu\nu},F^+_{\rho\lambda});\
(v^{(s)}_{\mu\nu},v^{(s)}_{\rho\lambda});\nonumber\\
& & (A,B)\equiv[F^+_{\mu\nu},v^{(s)}_{\rho\lambda}]_+.
\end{eqnarray}
The list (\ref{eq:list}) holds good for O$(q^4,\phi^{2n+1})$ terms with the
difference that there is an additional  factor of $i$ multiplying
the terms in $(i), (ii)$ and $(vi)$,
and the $i$ in $(iii), (iv)$ and $(v)$, is absent. The reason for
including $i$ only in some combinations of terms has to do with
imposing charge conjugation invariance 
along with other symmetries {\it directly within HBChPT} (See \cite{1n}). 
The terms of (\ref{eq:list}) and their analogs for O$(q^4,\phi^{2n+1})$
are not all independent since they can be related by a number of linear
relations: see next section (and \cite{1n} for O($q^3)$).

\section{Further Reduction due to Algebraic identities}

In this section, we discuss
further reduction in addition to the  ones obtained from (\ref{eq:phoffsh}).
The main result from \cite{1n} is that one need not consider
trace-dependent terms in SU(2) HBChPT if one assumes isospin conservation.
Given that isospin violation enters only via $\chi_\pm$, thus,
$\chi_\pm$-independent 
trace-dependent O$(q^4)$ terms can be eliminated in preference for 
$\chi_\pm$-independent trace-independent terms. We discuss reduction due to
algebraic identities in the  various categories of (\ref{eq:list}).
Some of the algebraic reductions require one to consider  more than one
category at a time, e.g., for  O$(q^4,\phi^{2n})$ terms,
the generalized 
Jacobi identities in (\ref{eq:oddanticoom}) require one to consider 
$(i), (i)(A\leftrightarrow B), (ii)$. 

One can show that (\cite{krause}):
\begin{eqnarray}
\label{eq:curvature}
& & [{\rm D}_\mu,{\rm D}_\nu]=
{1\over 4}[u_\mu, u_\nu]
-{i\over2}F^+_{\mu\nu}-iv^{(s)}_{\mu\nu},\\
& & [{\rm D}_\mu,u_\nu]-[{\rm D}_\nu,u_\mu]=F^-_{\mu\nu}.
\end{eqnarray}
The first relation, referred to as the curvature relation, will be
used extensively in conjunction with some generalized Jacobi identities
and Schouten's identity discussed below. As a consequence of (9), 
we will choose to always write $[\rm D_\mu,\rm D_\nu]$
in terms of $[u_\mu,u_\nu], F^+_{\mu\nu}, v^{(s)}_{\mu\nu}$. 
As a consequence of (10), we see that $F^-_{\mu\nu}$ 
can be eliminated in preference for $[{\rm D}_{[\mu},u_{\nu]}]$. However,
the relative coefficients of $[u_\mu,u_\nu], F^+_{\mu\nu}, v^{(s)}_{\mu\nu}$,
as well as the relative coefficients of $[{\rm D}_\mu,u_\nu], [{\rm D}_\nu,u_\mu]$
can be made arbitrary because each can be obtained from the nonrelativistic 
reduction of linearly independent terms. One interesting consequence of
(10) is that in the absence  of external fields, the commutator
of the covariant derivative and the axial-vector building
block is symmetric in the Lorentz indeces - something missed
in \cite{O4ext0}.

Further, another source of major reduction in number of terms is the
Schouten's identity:
\begin{equation}
\label{eq:Schid}
\epsilon^{\mu_1\mu_2\mu_3\mu_4}X^{\mu_5}+{\rm cyclic}=0,
\end{equation}
where $\rm X_\mu$ is any arbitrary (axial)vector.

It is because of  the curvature relation that one requires to
consider, e.g., some 
(4,0,0,0,0,0,0)-, (0,4,0,0,0,0,0)- (2,2,0,0,0,0,0)-,
(2,0,0,0,1,0,0)-, (2,0,0,0,0,0,1)-, (0,2,0,0,1,0,0)-,
(0,2,0,0,0,0,1)-, (0,0,0,0,2,0,0)-, (0,0,0,0,0,0,2)-
and (0,0,0,0,1,0,1)-type terms 
together in (\ref{eq:LCindoddanticoom}).
Due to Schouten's identity, e.g., the aforementioned 7-tuples 
with $\epsilon^{\mu\nu\rho\lambda}v_\rho{\rm S}_\lambda$ and
$\epsilon^{\mu\nu\rho\lambda}{\rm S}_\lambda v\cdot$ (where
$v_\rho$ is contracted with a building block) are required to be
considered together.

Finally, the O($q^2$) pion eom will be used for reduction
in the number of linearly independent terms:
\begin{equation}
\label{eq:pipieoom}
[{\rm D}_\mu,u^\mu]={i\over2}(\chi_-
-{1\over2}\langle\chi_-\rangle)\equiv{i\over2}\tilde\chi_-.
\end{equation}

\subsection{O($q^4,\phi^{2n})$ Terms}

In this subsection, we consider reduction in the number
of  independent O$(q^4,\phi^{2n})$ terms due to various algebraic
identities. The following are the algebraic identities
responsible  for reduction in number of O$(q^4,\phi^{2n})$
terms:
(\ref{eq:ABoddanticoom1}), (\ref{eq:ABoddanticoom2}), (\ref{eq:oddanticoom}),
(\ref{eq:evenanticoom}), 
(\ref{eq:evencoom}) (\ref{eq:umured}) and (\ref{eq:oddanticoom1}). 
For (\ref{eq:oddanticoom})
and (\ref{eq:evenanticoom}), there are two sets each of terms
(one $\epsilon^{\mu\nu\rho\lambda}$-dependent
and the other $\epsilon^{\mu\nu\rho\lambda}$-independent),
that need to be considered.

\underline{\underline{$p=q=0$ in (\ref{eq:prodbb}):
$(A,(B,(C,D))); ((A,B),(C,D))$}}

This includes $(i) - (iv)$ of (\ref{eq:list}). All
 terms in each of the first four types (of terms)
in (\ref{eq:list}) [$(i) - (iv)$] are linearly independent for 
unequal field operators A, B, C, D. 
However for (4,0,0,0,0,0,0), (0,4,0,0,0,0,0) and (2,2,0,0,0,0,0), L.C.-independent terms, one
needs to consider A=C, B=D in (i) in equation (\ref{eq:list}). Using
\begin{equation} 
\label{eq:ABoddanticoom1}
[A,[B,[A,B]_+]] = - [A,[B,[A,B]]_+] 
\end{equation}
only three of the four terms in (i) of equation (\ref{eq:list}), 
are linearly independent. Similarly, using
\begin{equation} 
\label{eq:ABoddanticoom2}
[[A,B],[A,B]_+] = - [[A,B]_+,[A,B]],
\end{equation}
only three of  the four terms in (ii) of equation (\ref{eq:list}), 
are linearly independent.

There are some reductions possible due to some generalized Jacobi identities by 
considering :  $(i), (i)(A\leftrightarrow B), 
(ii)$ of (\ref{eq:list})($\equiv\epsilon^{\mu\nu\rho\lambda}$ -independent terms), and 
$(iii),(iii)(A\leftrightarrow B),(iv)$ of (\ref{eq:list}) ($\equiv
\epsilon^{\mu\nu\rho\lambda}$-dependent terms).
The reason why one can not hope to get reductions by considering
any other pairs of types of terms in $(i) -  (iv)$ (in (\ref{eq:list})), 
is because one can get (linear) algebraic relationships only between those
terms which are (both) independent of (have) an overall factor of $i$. 

\underline{$(i), (i)(A\leftrightarrow B), (ii)$ of (\ref{eq:list})}

One can show the following 6 generalized Jacobi identities:
\begin{eqnarray}
\label{eq:oddanticoom}
& & [A,[B,[C,D]_+]] - [[A,B],[C,D]_+] = (i)(a)(A\leftrightarrow B)\nonumber\\
& & [A,[B,[C,D]_+]] - [[A,B]_+,[C,D]_+]_+ =-(i)(d)(A\leftrightarrow B)
\nonumber\\
& & [A,[B,[C,D]]_+] - [[A,B]_+,[C,D]] = -(i)(b)(A\leftrightarrow B)\nonumber\\
& & [A,[B,[C,D]]_+] - [[A,B],[C,D]]_+ = (i)(c)(A\leftrightarrow B)\nonumber\\
& & [A,[B,[C,D]]]_+ - [[A,B]_+,[C,D]] = -(i)(c)(A\leftrightarrow B)\nonumber\\
& & [A,[B,[C,D]_+]_+]_+ - [[A,B],[C,D]_+] = (i)(d)(A\leftrightarrow B).
\nonumber\\
\end{eqnarray}
Further, one can apply the following
to $(B,(C,D))$ contained in $(A,(B,(C,D)))$:
\begin{eqnarray} 
\label{eq:evencoom}
& & [B,[C,D]] - [[B,C],D] = [C,[B,D]]\nonumber\\
& & [B,[C,D]_+]_+ - [[B,C]_+,D]_+ = -[C,[B,D]]\nonumber\\
& & [B,[C,D]] - [[B,C]_+,D]_+ = -[C,[B,D]_+]_+,
\end{eqnarray}
and:
\begin{eqnarray}
\label{eq:oddanticoom1}
& & i[B,[C,D]]_+ - i[[B,C],D]_+ = i[C,[B,D]_+]\nonumber\\
& & i[B,[C,D]]_+ - i[[B,C]_+,D] = -i[C,[B,D]]_+\nonumber\\
& & i[B,[C,D]_+] - i[[B,C]_+,D] = -i[C,[B,D]_+].
\end{eqnarray}
The three 
identities in (\ref{eq:oddanticoom1}) 
are similar to the ones that occur in SUSY graded Lie algebra for
$B, D\equiv$ fermionic and $C\equiv$ bosonic fields, $B, C\equiv$ fermionic and 
$D\equiv$ bosonic fields,  
and $B,C,D\equiv$ fermionic fields, respectively. 

(1) Using (\ref{eq:oddanticoom}), (\ref{eq:curvature}), (\ref{eq:evencoom}) 
and (\ref{eq:oddanticoom1}),  one needs to consider
the following  
(4,0,0,0,0,0,0)-, (0,4,0,0,0,0,0)- (2,2,0,0,0,0,0)-,
(2,0,0,0,1,0,0)-, (2,0,0,0,0,0,1)-, (0,2,0,0,1,0,0)-,
(0,2,0,0,0,0,1)-, (0,0,0,0,2,0,0)-, (0,0,0,0,0,0,2)-
and (0,0,0,0,1,0,1)-type 
$\epsilon^{\mu\nu\rho\lambda}$-independent
terms together:
\begin{eqnarray}
\label{eq:LCindoddanticoom}
& & (v\cdot{\rm D},({\rm D}_\mu,(v\cdot{\rm D},{\rm D}^\mu))),\ 
({\rm D} _\mu,(v\cdot{\rm D},(v\cdot{\rm D},{\rm D}^\mu))),\ 
((v\cdot{\rm D},{\rm D}_\mu),(v\cdot{\rm D},{\rm D}^\mu))\nonumber\\
& & (v\cdot u,(u_\mu,(v\cdot u,u^\mu))),\ (u_\mu,(v\cdot u,(v\cdot u,u^\mu))),\
((v\cdot u,u_\mu),(v\cdot u, u^\mu)),\nonumber\\
& & (v\cdot{\rm D},({\rm D}_\mu,(v\cdot u,u^\mu))),\ 
({\rm D}_\mu,(v\cdot{\rm D},(u_\mu,v\cdot u))),\ 
((v\cdot{\rm D},{\rm D}_\mu),(v\cdot u,u^\mu))),\nonumber\\
& & 
(v\cdot u,(u_\mu,(v\cdot{\rm D},{\rm D}^\mu))),\ 
(u_\mu,(v\cdot u,(v\cdot{\rm D},{\rm D}^\mu))),\nonumber\\
& & (v\cdot{\rm D},(({\rm D}_\mu,v\cdot u),u^\mu)),\
(u^\mu,(({\rm D}_\mu,v\cdot u),v\cdot{\rm D}),\
((v\cdot{\rm D},u^\mu),({\rm D}_\mu,v\cdot u)),\nonumber\\
& & ({\rm D}_\mu,((v\cdot{\rm D},u^\mu),v\cdot u)),\
(v\cdot u,((v\cdot{\rm D},u^\mu),{\rm D}_\mu)),\nonumber\\
& & 
v^\nu(v\cdot{\rm D},({\rm D}_\mu,F^{+\ \mu\nu})),
\ v^\nu({\rm D}_\mu,(v\cdot{\rm D},F^{+\ \mu\nu})),
v^\nu((v\cdot{\rm D},{\rm D}_\mu),F^{+\ \mu\nu}),
\nonumber\\
& & 
v^\nu(v\cdot{\rm D},({\rm D}_\mu,v^{(s)\ \mu\nu})),\ 
v^\nu({\rm D}_\mu,(v\cdot{\rm D},v^{(s)\ \mu\nu})),
v^\nu((v\cdot{\rm D},{\rm D}_\mu),v^{(s)\ \mu\nu}),\ 
\nonumber\\
& & 
v^\nu(v\cdot u,(u_\mu,F^{+\ \mu\nu})),
\ v^\nu(u_\mu,(v\cdot u,F^{+\ \mu\nu})),
v^\nu((v\cdot u,u_\mu),F^{\mu\nu}),
\nonumber\\
& & 
v^\nu(v\cdot u,(u_\mu,v^{(s)\ \mu\nu})),\ 
v^\nu(u_\mu,(v\cdot u,v^{(s)\ \mu\nu})),
v^\nu((v\cdot u,u_\mu),v^{(s)\ \mu\nu}),\ 
\nonumber\\
& & 
v^\kappa F^+_{\kappa\mu}v_\sigma F^{+\ \sigma\mu},\
v^\kappa v^{(s)}_{\kappa\mu}v_\sigma F^{+\ \sigma\mu},\
v^\kappa v^{(s)}_{\kappa\mu}v_\sigma v^{(s)\ \sigma\mu}.
\nonumber\\
& & 
\end{eqnarray}
One needs to do a  careful counting of the total number of identities
that one can write down using (\ref{eq:oddanticoom}) and
(\ref{eq:curvature}), and the total number of terms in those
identities.

(2) Similarly, one will need
to consider 
(4,0,0,0,0,0,0)-, (0,4,0,0,0,0,0)- (2,2,0,0,0,0,0)-,
(2,0,0,0,1,0,0)-, (2,0,0,0,0,0,1)-, (0,2,0,0,1,0,0)-,
(0,2,0,0,0,0,1)-, (0,0,0,0,2,0,0)-, (0,0,0,0,0,0,2)-
and (0,0,0,0,1,0,1)-type 
terms together:
\begin{eqnarray}
\label{eq:LCindoddanticoom1}
& & ({\rm D}_\nu,({\rm D}_\mu,({\rm D}^\nu,{\rm D}^\mu))),\ 
(({\rm D}_\nu,{\rm D}_\mu),({\rm D}^\nu,{\rm D}^\mu))\nonumber\\
& & (u_\nu,(u_\mu,(u^\nu,u^\mu))),
\ ((u_\nu,u_\mu),(u^\nu, u^\mu)),\nonumber\\
& & ({\rm D}_\nu,({\rm D}_\mu,(u^\nu,u^\mu))),\ 
(({\rm D}_\nu,{\rm D}_\mu),(u^\nu,u^\mu)))\nonumber\\
& & (u_\nu,(u_\mu,({\rm D}^\nu,{\rm D}^\mu)))\nonumber\\
& & 
({\rm D}^\mu,({\rm D}^\nu,F^+_{\mu\nu})),\ 
([{\rm D} ^\mu,{\rm D}^\nu],F^+_{\mu\nu})),\nonumber\\ 
& & 
({\rm D}^\mu,({\rm D}^\nu,v^{(s)}_{\mu\nu})),\
([{\rm D}^\mu,{\rm D}^\nu],v^{(s)}_{\mu\nu})),
\nonumber\\
& & (u^\mu,(u^\nu,F^+_{\mu\nu})),\ 
([u^\mu,u^\nu],F^+_{\mu\nu})),\nonumber\\
& & 
(u^\mu,(u^\nu,v^{(s)}_{\mu\nu})),\
([u^\mu,u^\nu],v^{(s)}_{\mu\nu}))
\nonumber\\
& & (F^+_{\mu\nu}) ^2,\ v^{(s)}_{\mu\nu}F^{+\ \mu\nu},\
(v^{(s)}_{\mu\nu})^2
\end{eqnarray}

\underline{$(iii),(iii)(A\leftrightarrow B)$ and $(iv)$; 
(vi) of (\ref{eq:list})}

One can show the following generalized Jacobi identities to be true:
\begin{eqnarray}
\label{eq:evenanticoom}
 & & 
i[A,[B,[C,D]_+]_+] - i[[A,B]_+,[C,D]_+] = -(iii)(a)(A\leftrightarrow B)
\nonumber\\
& & i[A,[B,[C,D]_+]_+] - i[[A,B],[C,D]_+]_+ = (iii)(b)(A\leftrightarrow B)
\nonumber\\
& & i[A,[B,[C,D]_+]]_+ - i[[A,B]_+,[C,D]_+] = -(iii)(b)(A\leftrightarrow B)
\nonumber\\
& & i[A,[B,[C,D]]_+]_+ - i[[A,B],[C,D]] = (iii)(c)(A\leftrightarrow B)\nonumber\\
& & i[A,[B,[C,D]]_+]_+ - i[[A,B]_+,[C,D]_+] = -(iii)(d)(A\leftrightarrow B)
\nonumber\\
& & i[A,[B,[C,D]]] - i[[A,B],[C,D]] = (iii)(d)(A\leftrightarrow B)\nonumber\\
\end{eqnarray}
Again one can apply the  generalized Jacobi-iieuntities
(\ref{eq:evencoom}) and (\ref{eq:oddanticoom1})
to $(B,(C,D))$ contained in $(A,(B,(C,D)))$.

(3) 
The identities (\ref{eq:evenanticoom}), (\ref{eq:evencoom})
and (\ref{eq:oddanticoom1}) along with (9) 
require one  to consider the following category of  
$\epsilon^{\mu\nu\rho\lambda}$-dependent 
(4,0,0,0,0,0,0)-, (0,4,0,0,0,0,0)- (2,2,0,0,0,0,0)-,
(2,0,0,0,1,0,0)-, (2,0,0,0,0,0,1)-, (0,2,0,0,1,0,0)-,
(0,2,0,0,0,0,1)-, (0,0,0,0,2,0,0)-, (0,0,0,0,0,0,2)-
and (0,0,0,0,1,0,1)-type 
terms together:
\begin{eqnarray} 
\label{eq:LCevenanticoom}
& & i\epsilon^{\mu\nu\rho\lambda}v_\rho\Biggl[
({\rm D}_\mu,({\rm D}_\nu,({\rm D}_\lambda,{\rm S}\cdot{\rm D}))),\ 
(u_\mu,(u_\nu,(u_\lambda,{\rm S}\cdot u))),\nonumber\\
& & 
({\rm S}\cdot{\rm D},({\rm D}_\mu,({\rm D}_\nu,[{\rm D}_\nu,{\rm D}_\lambda]))\
({\rm D}_\mu,({\rm S}\cdot{\rm D},[{\rm D}_\nu,{\rm D}_\lambda])),\nonumber\\
& & ({\rm S}\cdot u,(u_\mu,([u_\nu,u_\lambda])),
\ (u_\mu,({\rm S}\cdot u,[u_\nu,u_\lambda])),\nonumber\\
& & ([u_\mu,u_\nu],({\rm D}_\lambda,{\rm S}\cdot{\rm D})),
\ (u_\mu,(u_\nu,({\rm D}_\lambda,{\rm S}\cdot{\rm D}))),\nonumber\\
& & ({\rm D}_\mu,({\rm S}\cdot{\rm D},[u_\nu,u_\lambda])),\
({\rm S}\cdot{\rm D},({\rm D}_\mu,[u_\nu,u_\lambda])),\nonumber\\
& & (u_\mu,({\rm S}\cdot u,[{\rm D}_\nu,{\rm D}_\lambda])),\ 
({\rm S}\cdot u,(u_\mu,[{\rm D}_\nu,{\rm D}_\lambda])),\nonumber\\
& & ({\rm D}_\mu,({\rm D}_\nu,(u_\lambda,{\rm S}\cdot u))),
\ ([{\rm D}_\mu,{\rm D}_\nu],(u_\lambda,{\rm S}\cdot u))\nonumber\\
& & ({\rm D}_\mu,(({\rm D}_\nu,
u_\lambda),{\rm S}\cdot u)),\ ({\rm S}\cdot u,(({\rm D}_\nu,
u_\lambda),{\rm D}_\mu));\nonumber\\ 
& & (({\rm D}_\mu,{\rm S}\cdot u),({\rm D}_\nu, u_\lambda));\nonumber\\  
& & ({\rm D}_\mu,(({\rm D}_\nu,{\rm S}\cdot u),u_\lambda)),\ 
(u_\lambda,(({\rm D}_\nu,{\rm S}\cdot u),{\rm D}_\mu)),\nonumber\\ 
& & 
i(F^+_{\mu\nu},({\rm D}_\lambda,{\rm S}\cdot{\rm D})),\
i({\rm D}_\lambda,({\rm S}\cdot{\rm D},F^+_{\mu\nu})),\
i({\rm S}\cdot{\rm D},({\rm D}_\lambda,F^+_{\mu\nu})),
\nonumber\\
& & 
i(v^{(s)}_{\mu\nu},({\rm D}_\lambda,{\rm S}\cdot{\rm D})),\
i({\rm D}_\lambda,({\rm S}\cdot{\rm D},v^{(s)}_{\mu\nu})),\
i({\rm S}\cdot{\rm D},({\rm D}_\lambda,v^{(s)}_{\mu\nu})),
\nonumber\\
& & 
i({\rm D}_\mu,({\rm D}_\nu,F^+_{\kappa\lambda}))S^\kappa,\ 
i([{\rm D}_\mu,{\rm D}_\nu],F^+_{\kappa\lambda})S^\kappa,\ 
i({\rm D}_\mu,({\rm D}_\nu,v^{(s)}_{\kappa\lambda}))S^\kappa,\nonumber\\
& & 
i(u_\mu,(u_\nu,F^+_{\kappa\lambda}))S^\kappa,\ 
i([u_\mu,u_\nu],F^+_{\kappa\lambda})S^\kappa,\ 
i(u_\mu,(u_\nu,v^{(s)}_{\kappa\lambda}))S^\kappa,\nonumber\\
& & i(F^+_{\mu\nu},(u_\lambda,{\rm S}\cdot u)),\
i(u_\lambda,({\rm S}\cdot u,F^+_{\mu\nu})),\
i({\rm S}\cdot u,(u_\lambda,F^+_{\mu\nu})),
\nonumber\\
& & 
i[u_\lambda,{\rm S}\cdot u]_+v^{(s)}_{\mu\nu},\
[F^+_{\mu\nu},F^+_{\kappa\lambda}]S^\kappa \Biggr]. 
\end{eqnarray}

Analogous to (\ref{eq:LCindoddanticoom}) and (\ref{eq:LCindoddanticoom1}),
one needs to do a  careful counting of the total number of identities
that one can write down using (\ref{eq:evenanticoom}),
and (\ref{eq:curvature}), 
and the total number of terms in those
identities.

A similar analysis can be carried out  for terms with $v\leftrightarrow\rm S$
in (\ref{eq:LCevenanticoom}). However, due to (\ref{eq:Schid}), this
set of terms has to be considered in conjunction with
(\ref{eq:LCevenanticoom1})
$(u_\kappa,{\rm D}^\kappa\rightarrow v\cdot u, v\cdot{\rm D})$

The identities in (\ref{eq:evencoom}) are also used in, e.g., 
$\epsilon^{\mu\nu\rho\lambda}$-dependent (1,1,0,1,0,0,0)-type terms.

(4) 
The identities (\ref{eq:evenanticoom}), (\ref{eq:evencoom}), 
(\ref{eq:oddanticoom1}) along with the curvature relation,
require one  to consider the following category of  
$\epsilon^{\mu\nu\rho\lambda}$-
dependent 
(4,0,0,0,0,0,0)-, (0,4,0,0,0,0,0)- (2,2,0,0,0,0,0)-,
(2,0,0,0,1,0,0)-, (2,0,0,0,0,0,1)-, (0,2,0,0,1,0,0)-,
(0,2,0,0,0,0,1)-, (0,0,0,0,2,0,0)-, (0,0,0,0,0,0,2)-
type terms together:
\begin{eqnarray} 
\label{eq:LCevenanticoom1}
& & i\epsilon^{\mu\nu\rho\lambda}v_\rho{\rm S}_\lambda\Biggl(
(u_\kappa,(u_\mu,(u^\kappa,u_\nu))),\ (u_\mu,(u_\kappa,(u^\kappa,u_\nu))),\
((u_\mu,u_\kappa),(u_\nu,u^\kappa)),\nonumber\\
& & (\rm D_\kappa,(\rm D_\mu,(\rm D^\kappa,\rm D_\nu))),\
((\rm D_\mu,(\rm D_\kappa,(\rm D^\kappa,\rm D_\nu))), \ 
((\rm D_\kappa,\rm D_\mu),(\rm D^\kappa,\rm D_\nu)),\nonumber\\
& & (({\rm D}_\kappa,({\rm D}_\mu,(u^\kappa,u_\nu))),\ 
(({\rm D}_\mu,({\rm D}_\kappa,(u^\kappa,u_\nu))),\ 
(({\rm D}_\kappa,{\rm D}_\mu),(u^\kappa,u_\nu)),\nonumber\\
& & (u_\mu,(u_\kappa,({\rm D}^\kappa,{\rm D}_\nu))),\
(u_\mu,(u_\kappa,({\rm D}^\kappa,{\rm D}_\nu))),\nonumber\\
&  &  
({\rm D}_\mu,(({\rm D}_\nu,u_\kappa),u^\kappa)),\ 
(({\rm D}_\mu,u^\kappa),({\rm D}_\nu,u^\kappa)),\nonumber\\
& & (u^\kappa,(({\rm D}_\mu,u^\kappa),{\rm D}_\nu)),\nonumber\\
& & i({\rm D}^\kappa,({\rm D}_\mu,F^+_{\kappa\nu})),\ 
i({\rm D}_\mu,({\rm D}^\kappa,F^+_{\kappa\nu})),\ 
i(({\rm D}^\kappa,{\rm D}_\mu),F^+_{\kappa\nu})\nonumber\\
& & i({\rm D}^\kappa,({\rm D}_\mu,v^{(s)}_{\kappa\nu})),\ 
i({\rm D}_\mu,({\rm D}^\kappa,v^{(s)}_{\kappa\nu})),\ 
i(({\rm D}^\kappa,{\rm D}_\mu),v^{(s)}_{\kappa\nu})\nonumber\\
& & i(u^\kappa,(u_\mu,F^+_{\kappa\nu})),\ 
i(u_\mu,(u^\kappa,F^+_{\kappa\nu})),\ 
i((u^\kappa,u_\mu),F^+_{\kappa\nu})\nonumber\\
& & i(u^\kappa,(u_\mu,v^{(s)}_{\kappa\nu})),\ 
i(u_\mu,(u^\kappa,v^{(s)}_{\kappa\nu})),\ 
i((u^\kappa,u_\mu),v^{(s)}_{\kappa\nu})\nonumber\\
& & i[F^{+\ \kappa}_\mu,F^+_{\kappa\nu}]. \Biggr).
\end{eqnarray} 

A similar  analysis can be carried out for terms with
$(u_\kappa,{\rm D}^\kappa)\rightarrow
(v\cdot u, v\cdot\rm D)$ in (\ref{eq:LCevenanticoom1}).

For (0,4,0,0,0,0,0)- and (1,3,0,0,0,0,0)-type
terms, writing $u_\mu=u_\mu^a\tau^a,\ [{\rm D}_\mu,u_\nu]=[{\rm D}_\mu,u_\nu]^a\tau^a$
(where $[{\rm D}_\mu,u_\nu]^a\equiv\partial_\mu u_\nu^a+i\epsilon^{abc}\Gamma_\mu^b u_\nu^c$), 
one will need to consider the following relations:
\begin{eqnarray}
\label{eq:umured}  
& (a) & [\tau^b,[\tau^c,\tau^d]_+]=[\tau^a,[\tau^b,[\tau^c,\tau^d]]_+]
=0;\ \nonumber\\ 
&  &  [[\tau^a,\tau^b],[\tau^c,\tau^d]_+]=0;\nonumber\\
&  & i[[\tau^a,\tau^b]_+,[\tau^c,\tau^d]_+]=0;\nonumber\\
& (b) & [u_\mu,u_\nu]_+^2-[u_\mu,u_\nu]^2=4u^4;\nonumber\\
& (c) & v\cdot u u_\mu v\cdot u u^\mu+h.c.
=-2u^2(v\cdot u)^2+[v\cdot u,u_\mu]_+^2;\nonumber\\
& (d) & i\epsilon^{\mu\nu\rho\lambda}[[u_\mu,u_\kappa],u_\nu,u^\kappa]_+]_+
=i\epsilon^{\mu\nu\rho\lambda}
\biggl(u^2[u_\mu,u_\nu]-u^\kappa[u_\mu,u_\nu]u_\kappa\biggr);
\nonumber\\
& (e) & i\epsilon^{\mu\nu\rho\lambda}[[u_\mu,v\cdot u],
[u_\nu,v\cdot u]_+]_+=
i\epsilon^{\mu\nu\rho\lambda}\biggl((v\cdot u)^2[u_\mu,u_\nu]-v\cdot
[u_\mu,u_\nu]v\cdot u\biggr).\nonumber\\
& & 
\end{eqnarray}

\underline{\underline{At least one of $k,p,q,t,u$ is 
$\neq0$ in (\ref{eq:prodbb}):$(A,(B,C))$}}

This includes $(v)$ and $(vi)$ of (\ref{eq:list}).

\underline{($v$) of (\ref{eq:list})} 

The identities
in (\ref{eq:oddanticoom1}) are used in, e.g., 
$\epsilon^{\mu\nu\rho\lambda}$-independent (1,1,0,1,0,0,0)-type terms. When
applying (\ref{eq:oddanticoom1}) to (2,0,1,0,0,0,0), because
of (\ref{eq:curvature}), one will need to consider the following
terms together:
\begin{eqnarray}
\label{eq:epsDuchi+}
& & i\epsilon^{\mu\nu\rho\lambda}v_\rho{\rm S}_\lambda\biggl(
({\rm D}_\mu,({\rm D}_\nu,\chi_+)),\
([{\rm D}_\mu,{\rm D}_\nu],\chi_+);\nonumber\\
& & (u_\mu,(u_\nu,\chi_+)),\ ([u_\mu,u_\nu],\chi_+),\
i(F^+_{\mu\nu},\chi_+),\ iv^{(s)}_{\mu\nu},\chi_+)
\biggr).
\end{eqnarray}

\subsection{O$(q^4,\phi^{2n+1}$) Terms}

In this subsection, we consider the reduction in the 
number of O$(q^4,\phi^{2n+1})$ terms because of algebraic identities.
The discussion in this subsection will be much briefer than the preceding
(subsection).

\underline{$(i)-(iv)$ of (\ref{eq:list})$^\prime$}

(1) The identities (\ref{eq:oddanticoom})
are the same for O$(q^4,\phi^{2n+1})$
except for an overall factor of $i$. We will denote the analogue
of (\ref{eq:oddanticoom}) for O$(q^4,\phi^{2n+1})$ terms 
as (\ref{eq:oddanticoom})$^\prime$.\footnote{Similarly, the analogs of
(\ref{eq:list}), (\ref{eq:evenanticoom}), (\ref{eq:oddanticoom1}) 
and (\ref{eq:evencoom})  will be denoted
by (\ref{eq:list})$^\prime$, 
(\ref{eq:evenanticoom})$^\prime$, (\ref{eq:oddanticoom1})$^\prime$ 
and (\ref{eq:evencoom})$^\prime$.}
Using it together with 
and (\ref{eq:curvature}), one sees that
one needs to consider the following set of terms together:
\begin{eqnarray}
\label{eq:oddset1}
& (a) & i(v\cdot{\rm D},({\rm D}_\mu,({\rm D}^\mu,{\rm S}\cdot u))),\
i({\rm D}_\mu,(v\cdot{\rm D},({\rm D}^\mu,{\rm S}\cdot u))),\ 
i((v\cdot{\rm D},{\rm D}_\mu),({\rm D}^\mu,{\rm S}\cdot u)),\nonumber\\
& & i({\rm S}\cdot u,({\rm D}_\mu,({\rm D}^\mu,v\cdot{\rm D}))),\
i({\rm D}_\mu,({\rm S}\cdot u,({\rm D}^\mu,v\cdot{\rm D})));\nonumber\\
& & ({\rm D}^\mu,({\rm S}\cdot u,F^+_{\mu\nu}))v^\nu,
({\rm S}\cdot u,({\rm D}^\mu,F^+_{\mu\nu}))v^\nu,
(({\rm D}^\mu,{\rm S}\cdot u),F^+_{\mu\nu})v^\nu,\nonumber\\
& & ({\rm D}^\mu,({\rm S}\cdot u,v^{(s)}_{\mu\nu}))v^\nu,
({\rm S}\cdot u,({\rm D}^\mu,v^{(s)}_{\mu\nu}))v^\nu,
(({\rm D}^\mu,{\rm S}\cdot u),v^{(s)}_{\mu\nu})v^\nu;\nonumber\\
&  & i(v\cdot u,(u_\mu,({\rm S}\cdot u,{\rm D}^\mu))),\
i(u_\mu,(v\cdot u,({\rm S}\cdot u,{\rm D}^\mu))),\
i((v\cdot u,u_\mu),({\rm S}\cdot u,{\rm D}^\mu)),\nonumber\\
& & i({\rm D}_\mu,({\rm S}\cdot u,(u^\mu,v\cdot u)))\
i({\rm S}\cdot u,({\rm D}_\mu,(u^\mu,v\cdot u)));\nonumber\\
& & i({\rm D}_\mu,({\rm D}^\mu,(v\cdot{\rm D},{\rm S}\cdot u))),\
i({\rm D}^2,(v\cdot{\rm D},{\rm S}\cdot u)),\nonumber\\
& & i({\rm S}\cdot u,(v\cdot{\rm D},{\rm D}^2)),\ 
i(v\cdot{\rm D},({\rm S}\cdot u,{\rm D}^2)).\nonumber\\
& (b) & i(v\cdot{\rm D},({\rm D}_\mu,({\rm S}\cdot{\rm D},u^\mu))),\
i({\rm D}_\mu,(v\cdot{\rm D},({\rm S}\cdot{\rm D},u^\mu))),\
i((v\cdot{\rm D},{\rm D}_\mu),({\rm S}\cdot{\rm D},u^\mu)),\nonumber\\
& & i(u_\mu,({\rm S}\cdot{\rm D},({\rm D}^\mu,v\cdot{\rm D})))\
i({\rm S}\cdot{\rm D},(u_\mu,({\rm D}^\mu,v\cdot{\rm D})));\nonumber\\
& & (u^\mu,({\rm S}\cdot {\rm D},F^+_{\mu\nu}))v^\nu,
({\rm S}\cdot {\rm D},(u^\mu,F^+_{\mu\nu}))v^\nu,
((u^\mu,{\rm S}\cdot{\rm D}),F^+_{\mu\nu})v^\nu,\nonumber\\
& & (u^\mu,({\rm S}\cdot {\rm D},v^{(s)}_{\mu\nu}))v^\nu,
({\rm S}\cdot {\rm D},(u^\mu,v^{(s)}_{\mu\nu}))v^\nu,
(u^\mu,{\rm S}\cdot {\rm D}),v^{(s)}_{\mu\nu})v^\nu\nonumber\\
& & i(v\cdot u,(u_\mu,(u^\mu,{\rm S}\cdot {\rm D}))),\
i(u_\mu,(v\cdot u,(u^\mu,{\rm S}\cdot {\rm D}))),\ 
i((v\cdot u,u_\mu),(u^\mu,{\rm S}\cdot{\rm D})),\nonumber\\
& & i({\rm S}\cdot {\rm D},(u_\mu,(u^\mu,v\cdot u))),\
i(u_\mu,({\rm S}\cdot {\rm D},(u^\mu,v\cdot u)))\nonumber\\
& & i(u^\mu,(u_\mu,(v\cdot u,{\rm D}^\mu))),\
i({\rm S}\cdot u,(u_\mu,v\cdot u,{\rm D}^\mu)),\nonumber\\
& & i((u_\mu,{\rm S}\cdot u),(v\cdot u,{\rm D}^\mu)),\
i({\rm D}_\mu,(v\cdot u,(u^\mu,{\rm S}\cdot u))),
\nonumber\\
& & i(v\cdot u,({\rm D}_\mu,(u^\mu,{\rm S}\cdot u)));\nonumber\\
& & i(u^\mu,(u_\mu,(v\cdot u,{\rm S}\cdot{\rm D}))),\ 
i(u^2,(v\cdot u,{\rm S}\cdot{\rm D})),\nonumber\\
& & i({\rm S}\cdot{\rm D},(v\cdot u,u^2)),
\ i(v\cdot u,({\rm S}\cdot{\rm D},u^2));
\nonumber\\
& (c) & 
i(v\cdot{\rm D},({\rm S}\cdot{\rm D},({\rm D}^\mu,u_\mu))),\
i({\rm S}\cdot{\rm D},(v\cdot{\rm D},({\rm D}^\mu,u_\mu))),\
i((v\cdot{\rm D},{\rm S}\cdot{\rm D}),({\rm D}^\mu,u_\mu)),\nonumber\\
& & i(u_\mu,({\rm D}^\mu,(v\cdot{\rm D},{\rm S}\cdot{\rm D}))),\
i({\rm D}_\mu,(u^\mu,(v\cdot{\rm D}{\rm S}\cdot{\rm D})));\nonumber\\
& & 
v^{[\rho}{\rm S}^{\lambda]}({\rm D}_\mu,(u^\mu,F^+_{\rho\lambda})),
v^{[\rho}{\rm S}^{\lambda]}(u^\mu,({\rm D}_\mu,F^+_{\rho\lambda})),
v^{[\rho}{\rm S}^{\lambda]}(({\rm D}_\mu,u^\mu),F^+_{\rho\lambda})\nonumber\\
& & 
v^{[\rho}{\rm S}^{\lambda]}({\rm D}_\mu,(u^\mu,v^{(s)}_{\rho\lambda})),
v^{[\rho}{\rm S}^{\lambda]}(u^\mu,({\rm D}_\mu, v^{(s)}_{\rho\lambda})),
v^{[\rho}{\rm S}^{\lambda]}(({\rm D}_\mu,u^\mu),v^{(s)}_{\rho\lambda})
\nonumber\\
& & 
i(v\cdot  u,({\rm S}\cdot u,(u^\mu,{\rm D}_\mu))),\
i({\rm S}\cdot u,(v\cdot u,({\rm D}^\mu,u_\mu))),\
i((v\cdot u,{\rm S}\cdot u),({\rm D}^\mu,u_\mu)),\nonumber\\
& & i({\rm D}_\mu,(u^\mu,(v\cdot u,{\rm S}\cdot u))),\
i(u_\mu,({\rm D}^\mu,(v\cdot u,{\rm S}\cdot u)));\nonumber\\
& (d) & 
i({\rm D}_\mu,({\rm S}\cdot{\rm D},(v\cdot{\rm D},u^\mu))),\
i({\rm S}\cdot{\rm D},({\rm D}_\mu,(v\cdot{\rm D},u^\mu))),\
i(({\rm S}\cdot{\rm D},{\rm D}_\mu),(v\cdot{\rm D},u^\mu)),\nonumber\\
& & i(u_\mu,(v\cdot{\rm D},({\rm D}^\mu,{\rm S}\cdot{\rm D}))),\
i(v\cdot{\rm D},(u_\mu,({\rm D}^\mu,{\rm S}\cdot{\rm D}))),\nonumber\\
& & (u^\mu,(v\cdot{\rm D},F^+_{\mu\nu})){\rm S}^\nu,
(v\cdot{\rm D},(u^\mu,F^+_{\mu\nu})){\rm S}^\nu,\
((u^\mu,v\cdot{\rm D}),F^+_{\mu\nu}){\rm S}^\nu,\nonumber\\
& & (u^\mu,(v\cdot{\rm D},v^{(s)}_{\mu\nu})){\rm S}^\nu,
(v\cdot{\rm D},(u^\mu,v^{(s)}_{\mu\nu})){\rm S}^\nu,\
((u^\mu,v\cdot{\rm D}),v^{(s)}_{\mu\nu}){\rm S}^\nu,\nonumber\\
& & i(u_\mu,({\rm S}\cdot u,(u^\mu,v\cdot {\rm D}))),\
i({\rm S}\cdot u,(u_\mu,(u^\mu,v\cdot {\rm D}))),\
i((u_\mu,{\rm S}\cdot u),(u^\mu,v\cdot {\rm D}))),\nonumber\\
& & i(v\cdot {\rm D},(u_\mu,(u^\mu,{\rm S}\cdot u)),\ 
i(u_\mu,(v\cdot {\rm D},(u^\mu,{\rm S}\cdot u));\nonumber\\
& & i(u_\mu,(u^\mu,({\rm S}\cdot u,v\cdot{\rm D}))),\
i(u^2,({\rm S}\cdot u,v\cdot{\rm D})),\nonumber\\
& & i(v\cdot{\rm D},({\rm S}\cdot u,u^2)),\ 
i({\rm S}\cdot u,(v\cdot{\rm D},u^2));\nonumber\\
& (e) &
i({\rm D}_\mu,({\rm S}\cdot{\rm D},({\rm D}^\mu,v\cdot u))),\
i({\rm S}\cdot{\rm D},({\rm D}_\mu,({\rm D}^\mu,v\cdot u))),\
i(({\rm D}_\mu,{\rm S}\cdot{\rm D}),({\rm D}^\mu,v\cdot u))),\nonumber\\
& & i(v\cdot u,({\rm D}_\mu,({\rm D}^\mu,{\rm S}\cdot{\rm D})),\ 
i({\rm D}_\mu,(v\cdot u,({\rm D}^\mu,{\rm S}\cdot{\rm D}));\nonumber\\
& & (v\cdot u,({\rm D}^\mu,F^+_{\mu\nu})){\rm S}^\nu,
({\rm D}^\mu,(v\cdot u,F^+_{\mu\nu})){\rm S}^\nu,\
((v\cdot u,{\rm D}^\mu),F^+_{\mu\nu}){\rm S}^\nu,\nonumber\\
& & (v\cdot u,({\rm D}^\mu,v^{(s)}_{\mu\nu})){\rm S}^\nu,
({\rm D}^\mu,(v\cdot u,v^{(s)}_{\mu\nu})){\rm S}^\nu,\
((v\cdot u,{\rm D}^\mu),v^{(s)}_{\mu\nu}){\rm S}^\nu,\nonumber\\
& & i(u_\mu,({\rm S}\cdot u,(v\cdot u,{\rm D}^\mu))),\
i({\rm S}\cdot u,(u_\mu,(v\cdot u,{\rm D}^\mu))),\
i(({\rm S}\cdot u,u_\mu),(v\cdot u,{\rm D}^\mu)),\nonumber\\
& & i({\rm D}_\mu,(v\cdot u,(u^\mu,{\rm S}\cdot u))),\
i(v\cdot u,({\rm D}_\mu,(u^\mu,{\rm S}\cdot u)));\nonumber\\
& & i({\rm D}_\mu,({\rm D}^\mu,({\rm S}\cdot{\rm D},v\cdot u))),\
i({\rm D}^2,({\rm S}\cdot{\rm D},v\cdot u)),\nonumber\\
& & i(v\cdot u,({\rm S}\cdot{\rm D},{\rm D}^2)),\ 
i({\rm S}\cdot{\rm D},(v\cdot u,{\rm D}^2)).
\end{eqnarray}
Using (\ref{eq:evencoom})$^\prime$ and (\ref{eq:oddanticoom1})$^\prime$, one
needs to consider $(a)-(e)$ together.

Similarly, using (\ref{eq:oddanticoom})$\prime$ and 
(\ref{eq:curvature})$^\prime$,
one needs to consider simultaneously
\begin{eqnarray}
& (a) & 
i(v\cdot{\rm D},({\rm S}\cdot{\rm D},(v\cdot {\rm D},v\cdot u))),\
i({\rm S}\cdot{\rm D},(v\cdot{\rm D},(v\cdot {\rm D},v\cdot u))),\
i((v\cdot{\rm D},{\rm S}\cdot{\rm D}),(v\cdot {\rm D},v\cdot u))),\nonumber\\
& & i(v\cdot u,(v\cdot{\rm D},(v\cdot{\rm D},{\rm S}\cdot{\rm D})),\
i(v\cdot {\rm D},(v\cdot u,(v\cdot{\rm D},{\rm S}\cdot{\rm D}));\nonumber\\
& & v^{[\rho}{\rm S}^{\lambda]}(v\cdot{\rm D},(v\cdot u,F^+_{\rho\lambda})),
v^{[\rho}{\rm S}^{\lambda]}(v\cdot u,(v\cdot{\rm D},F^+_{\rho\lambda})),
v^{[\rho}{\rm S}^{\lambda]}((v\cdot{\rm D},v\cdot u),
F^+_{\rho\lambda});\nonumber\\
& & 
v^{[\rho}{\rm S}^{\lambda]}(v\cdot {\rm D},(v\cdot u,v^{(s)}_{\rho\lambda})),
v^{[\rho}{\rm S}^{\lambda]}(v\cdot u,(v\cdot{\rm D}, v^{(s)}_{\rho\lambda})),
v^{[\rho}{\rm S}^{\lambda]}((v\cdot{\rm D},v\cdot u),v^{(s)}_{\rho\lambda});
\nonumber\\
& & 
i(v\cdot u,({\rm S}\cdot u,(v\cdot u,v\cdot {\rm D}))),\
i({\rm S}\cdot u,(v\cdot u,(v\cdot u,v\cdot {\rm D}))),\
i(({\rm S}\cdot u,v\cdot u),(v\cdot u,v\cdot {\rm D})),\nonumber\\
& & i(v\cdot u,(v\cdot{\rm D},(v\cdot u,{\rm S}\cdot u)),\
i(v\cdot {\rm D},(v\cdot u,(v\cdot u,{\rm S}\cdot u));\nonumber\\
&  & i(v\cdot u,({\rm S}\cdot{\rm D},(v\cdot{\rm D})^2)),\ 
i({\rm S}\cdot{\rm D},(v\cdot u,(v\cdot{\rm D})^2)),\nonumber\\
& & i((v\cdot{\rm D})^2,({\rm S}\cdot{\rm D},v\cdot u)),\ 
i(v\cdot{\rm D},(v\cdot{\rm D},({\rm S}\cdot{\rm D},v\cdot u)));\nonumber\\
& & i(v\cdot{\rm D},{\rm S}\cdot u,(v\cdot u)^2)),\
i({\rm S}\cdot u,(v\cdot{\rm D},(v\cdot u)^2)),\nonumber\\
& & i((v\cdot u)^2,({\rm S}\cdot u,v\cdot{\rm D})),\
i(v\cdot u,(v\cdot u,({\rm S}\cdot u,v\cdot{\rm D}));\nonumber\\
& (b) & i(v\cdot{\rm D},(v\cdot{\rm D},(v\cdot{\rm D},{\rm S}\cdot u))),\
i((v\cdot{\rm D})^2,(v\cdot{\rm D},{\rm S}\cdot u)),\nonumber\\
& & i[{\rm S}\cdot u,(v\cdot{\rm D})^3]_+,
\ i(v\cdot{\rm D},({\rm S}\cdot u,(v\cdot{\rm D})^2)).
\end{eqnarray}
Using (\ref{eq:evencoom})$^\prime$ and (\ref{eq:oddanticoom1})$^\prime$,
one needs to consider $(a)$ and $(b)$ together.

(2) Using (\ref{eq:evenanticoom})$^\prime$, (\ref{eq:evencoom})$^\prime$,
(\ref{eq:oddanticoom1})$^\prime$, (\ref{eq:curvature}) and 
(\ref{eq:Schid}),
one sees that one has  to consider the following sets of terms together:
\begin{eqnarray}
\label{eq:oddset11}
& (a) & \epsilon^{\mu\nu\rho\lambda}\biggl(
({\rm D}_\mu,({\rm D}_\nu,({\rm D}_\rho,u_\lambda))),\
([{\rm D}_\mu,{\rm D}_\nu],({\rm D}_\rho,u_\lambda))),\nonumber\\
& & (u_\mu,({\rm D}_\nu,[{\rm D}_\rho,{\rm D}_\lambda])),\
({\rm D}_\mu,(u_\nu,[{\rm D}_\rho,{\rm D}_\lambda]));\nonumber\\
& & 
i({\rm D}_\mu,(u_\nu,F^+_{\rho\lambda})), 
i(u_\nu,({\rm D}_\mu,F^+_{\rho\lambda})), 
i((u_\nu,{\rm D}_\mu),F^+_{\rho\lambda});\nonumber\\
& & i(u_\mu,({\rm D},v^{(s)}_{\rho\lambda})), 
i({\rm D}_\mu,u_\nu)v^{(s)}_{\rho\lambda};
\nonumber\\
& & (u_\mu,(u_\nu,(u_\rho,{\rm D}_\lambda))),\
([u_\mu,u_\nu],(u_\rho,{\rm D}_\lambda)),\nonumber\\
& & ({\rm D}_\mu,(u_\nu,[u_\rho, u_\lambda])),\
(u_\mu,({\rm D}_\nu,[u_\rho,u_\lambda]))\biggr);\nonumber\\
& (b) & \epsilon^{\mu\nu\rho\lambda}v_\rho\biggl(
(v\cdot{\rm D},({\rm D}_\mu,({\rm D}_\nu,u_\lambda))),\
((v\cdot{\rm D},{\rm D}_\mu),({\rm D}_\nu,u_\lambda))),\
({\rm D}_\mu,(v\cdot{\rm D},({\rm D}_\nu,u_\lambda))),\nonumber\\
& & (u_\mu,({\rm D}_\nu,(v\cdot{\rm D},{\rm D}_\lambda))),\
({\rm D}_\mu,(u_\nu,(v\cdot{\rm D},{\rm D}_\lambda)));\nonumber\\
& & (v\cdot u,(u_\mu,(u_\nu,{\rm D}_\lambda))),\
(u_\mu,(v\cdot u,(u_\nu,{\rm D}_\lambda))),\
((u_\mu,v\cdot u),(u_\nu,{\rm D}_\lambda)),\nonumber\\
& & ({\rm D}_\mu,(u_\nu,(v\cdot u,u_\lambda))),\
(u_\mu,({\rm D}_\nu,(v\cdot u,u_\lambda)))\biggr),\nonumber\\
& & i({\rm D}_\mu,(u_\nu,F^+_{\kappa\lambda}))v^\kappa, 
i(u_\nu,({\rm D}_\mu,F^+_{\kappa\lambda}))v^\kappa, 
i((u_\nu,{\rm D}_\mu),F^+_{\kappa\lambda})v^\kappa;\nonumber\\
& & i(u_\mu,({\rm D},v^{(s)}_{\kappa\lambda}))v^\kappa, 
i({\rm D}_\mu,u_\nu)v^{(s)}_{\kappa\lambda}v^\kappa;\nonumber\\
& (c) & \epsilon^{\mu\nu\rho\lambda}v_\rho\biggl(
({\rm D}_\mu,({\rm D}_\nu,(v\cdot{\rm D},u_\lambda))),
([{\rm D}_\mu,{\rm D}_\nu],(v\cdot{\rm D},u_\lambda));
\nonumber\\
& & 
(v\cdot{\rm D},(u_\lambda,[{\rm D}_\mu,{\rm D}_\nu])),
(u_\mu,(v\cdot{\rm D},[{\rm D}_\mu,{\rm D}_\nu]));\nonumber\\
& & 
(u_\mu,(u_\nu,(v\cdot{\rm D},u_\lambda))),
([u_\mu,u_\nu],(v\cdot{\rm D},u_\lambda)));\nonumber\\
& & (u_\lambda,(v\cdot{\rm D},[u_\mu,u_\nu])),
(v\cdot{\rm D},(u_\lambda,[u_\mu,u_\nu]))
\nonumber\\
& & 
(v\cdot{\rm D},(u_\lambda,F^+_{\mu\nu})),
(u_\lambda, (v\cdot{\rm D}, F^+_{\mu\nu})),
((u_\lambda, v\cdot{\rm D}), F^+_{\mu\nu}));\nonumber\\
& & 
(v\cdot{\rm D},(u_\lambda,v^{(s)}_{\mu\nu})),
(u_\lambda, (v\cdot{\rm D}, v^{(s)}_{\mu\nu})),
((u_\lambda, v\cdot{\rm D}), v^{(s)}_{\mu\nu}));\nonumber\\
& (d) & \epsilon^{\mu\nu\rho\lambda}v_\rho\biggl(
({\rm D}_\mu,({\rm D}_\nu,({\rm D}_\lambda,v\cdot u))),
([{\rm D}_\mu,{\rm D}_\nu],(v\cdot u,{\rm D}_\lambda));
\nonumber\\
& & 
({\rm D}_\lambda,(v\cdot u,[{\rm D}_\mu,{\rm D}_\nu])),
(v\cdot u,({\rm D}_\lambda,[{\rm D}_\mu,{\rm D}_\nu]));\nonumber\\
& & 
(u_\mu,(u_\nu,(v\cdot u,{\rm D}_\lambda))),
([u_\mu,u_\nu],(v\cdot u,{\rm D}_\lambda)));\nonumber\\
& & (v\cdot u,({\rm D}_\lambda,[u_\mu,u_\nu])),
({\rm D}_\lambda,(v\cdot u,[u_\mu,u_\nu]));
\nonumber\\
& & 
({\rm D}_\lambda,(v\cdot u,F^+_{\mu\nu})),
(v\cdot u, ({\rm D}_\lambda, F^+_{\mu\nu})),
((v\cdot u, {\rm D}_\lambda), F^+_{\mu\nu}));\nonumber\\
& & 
({\rm D}_\lambda,(v\cdot u,v^{(s)}_{\mu\nu})),
(v\cdot u, ({\rm D}_\lambda, v^{(s)}_{\mu\nu})),
((v\cdot u, {\rm D}_\lambda), v^{(s)}_{\mu\nu})).
\end{eqnarray}

\underline{\underline{At least one of $k,p,q,t,u$ is 
$\neq0$ in (\ref{eq:prodbb}):$(A,(B,C))$}}

\underline{$(v)$ of (\ref{eq:list})$^\prime$}

Using (\ref{eq:curvature}) and (\ref{eq:oddanticoom1})$^\prime$, 
one sees that one will have to consider the following set of terms  together:
\begin{eqnarray}
\label{eq:oddset15}
& & 
({\rm S}\cdot{\rm D},(v\cdot{\rm D},\chi_-)),\
(v\cdot{\rm D},({\rm S}\cdot{\rm D},\chi_-)),\
((v\cdot{\rm D},{\rm S}\cdot{\rm D}),\chi_-);\nonumber\\
& & 
i[F^+_{\mu\nu},\chi_-]_+v^{[\mu}{\rm S}^{\nu]},
iv^{(s)}_{\mu\nu}\chi_-v^{[\mu}{\rm S}^{\nu]};\nonumber\\
& & (v\cdot u,({\rm S}\cdot u,\chi_-)),\ 
({\rm S}\cdot u,(v\cdot u,\chi_-)),\ 
(({\rm S}\cdot u,v\cdot u),\chi_-).
\end{eqnarray}

\underline{$(vi)$ of (\ref{eq:list})$^\prime$}

Using (\ref{eq:evencoom})$^\prime$, one sees that one will have
to consider the following set of terms together:
\begin{equation}
\label{eq:oddset16}
i({\rm S}\cdot{\rm D},(v\cdot u,\chi_+)),
\ i(v\cdot u,({\rm S}\cdot {\rm D},\chi_+)),\
i(({\rm S}\cdot{\rm D},v\cdot u),\chi_+);
v\leftrightarrow{\rm S}.
\end{equation}

Note that  because of  parity constraints and
the algebra of the $\rm S_\mu$s (See \cite{1n}), 
there are no Levi Civita-dependent 
(2,0,0,1,0,0,0)-, (0,2,0,1,0,0,0)- and (1,1,1,0,0,0,0)-type terms.

\subsection{Isospin Violation}
As noted in Section 3, isospin violation enters via $\chi_\pm$, we thus
need to reconsider the $(m,n,p,q,t,u,k)$ with $p$ or $q\neq0$. We will
also have to include trace-dependent terms for these type   
of terms. The following identities are used in arriving
at terms in Table 2:
\begin{eqnarray}
\label{eq:miscidisovioml}
& & u^\mu\chi_+u_\mu+u^2\chi_++{\rm h.c.}=2u^2\langle\chi_+\rangle  
+u^\mu\langle[u_\mu,\chi_+]_+\rangle,
\nonumber\\
& &
v\cdot u\chi_+v\cdot u+u^2\chi_++{\rm h.c.}
= 2(v\cdot u)^2\langle\chi_+\rangle
+v\cdot u\langle[v\cdot u,\chi_+]_+\rangle,\nonumber\\
& &[\chi_\pm,\chi_\pm]_+=2\langle\chi_\pm\rangle\chi_\pm+{1\over2} 
\langle[\chi_\pm,\chi_\pm]_+\rangle-{1\over 2}\langle\chi_\pm\rangle^2,
\nonumber\\
& &
[F^+_{\mu\nu},\chi_\pm]_+={1\over2}[F^+_{\mu\nu},\langle\chi_\pm\rangle]_+
+{1\over2}\langle[F^+_{\mu\nu},\chi_\pm]_+\rangle,\nonumber\\
& &
i[v\cdot{\rm D},[{\rm S}\cdot u,\chi_+]_+]_+
={i\over2}[v\cdot{\rm D},[{\rm S}\cdot u,\langle\chi_+\rangle]_+]_+
+{i\over2}[v\cdot{\rm D},\langle[{\rm S}\cdot
u,\chi_+]_+\rangle,\nonumber\\
& &
i[v\cdot{\rm D},[{\rm S}\cdot u,\chi_+]_+]_+
={i\over2}[v\cdot{\rm D},[{\rm S}\cdot u,\langle\chi_+\rangle]_+]_+
+{i\over2}[v\cdot{\rm D},\langle[{\rm S}\cdot
u,\chi_+]_+\rangle,\nonumber\\
& & +{i\over2}[v\cdot{\rm D},\langle[{\rm S}\cdot
u,\chi_+]_+\rangle,\nonumber\\
& & \epsilon^{\mu\nu\rho\lambda}v_\rho{\rm S}_\lambda
\biggl([{\rm D}_\mu,[u_\nu,\chi_-]_+]_+
=[{\rm D}_\mu,[u_\nu,\langle\chi_-\rangle]_+]_+
+{1\over2}[{\rm D}_\mu,\langle[u_\nu,\chi_-]_+\rangle]_+\biggr),\nonumber\\
& & i\epsilon^{\mu\nu\rho\lambda}v_\rho{\rm S}_\lambda
\biggl([[u_\mu,u_\nu],\chi_+]_+=[[u_\mu,u_\nu],\langle\chi_+\rangle]_+
+{1\over2}[[u_\mu,u_\nu],^H\chi_+]_+\rangle\biggr),\nonumber\\
& &
i[{\rm D}_\mu,[u^\mu,\chi_-]_+]={i\over2}\langle
[{\rm D}_\mu,[u^\mu,\chi_-]_+]\rangle
+{i\over2}[{\rm D}_\mu,[u^\mu,\langle\chi_-\rangle]_+],\nonumber\\
& &
i[[{\rm D}_\mu,u^\mu],\chi_-]_+={i\over2}\langle
[[{\rm D}_\mu,u^\mu],\chi_-]_+\rangle
+{i\over2}[[{\rm D}_\mu,u^\mu],\langle\chi_-\rangle]_+,\nonumber\\
& & i[u_\mu,[{\rm D}^\mu,\chi_-]]_+={i\over2}\langle[u_\mu,[{\rm D}^\mu,
\chi_-]]_+\rangle+{i\over2}[u_\mu,[{\rm D}^\mu,\langle\chi_-\rangle]_+. 
\end{eqnarray}

\section{The Lists of Independent Terms in ${\cal L}_{\rm HBChPT}$
(off-shell nucleons)}

In this section, using (\ref{eq:phoffsh}), and the algebraic reductions
of Section 3, we list  all possible
${\cal A}$-type terms of O$(q^4,\phi^{2n})$, 
and O$(q^4,\phi^{2n+1})$ in Tables
1 and 2, that are allowed by (\ref{eq:phoffsh})
and have not been eliminated in Section 3.
As noted in Section 2 (and \cite{1n}), for off-shell nucleons,
$\gamma^0{\cal B}^\dagger\gamma^0{\cal C}^{-1}{\cal B}\in {\cal A}$.
Hence, it is sufficient to list only ${\cal A}$-type terms (for off-shell
nucleons).

Using the algebraic identities
of Section 3, if we end up with $m$ independent identities in $n(>m)$ terms,
then we  can take $(n-m)$ linearly independent terms. 
Even though the phase rule (\ref{eq:phoffsh}) and linear independence of
terms are
sufficient for listing terms in the O$(q^4$) HBChPT Lagrangian for off-shell
nucleons, however, if for a given choice of terms and 
group of terms in (\ref{eq:list}), we find similar group of terms
in \cite{mms}, then while  listing the $(n-m)$ terms,
preference is  given to including terms that also
figure in Table 1 of \cite{mms}. The reason for doing
the same is that this allows for an easy identification of 
the finite terms, given
that the divergent (counter) terms have been worked out in \cite{mms}.

In tables 1 (that one gets if one assumes isospin symmetry) and 2 (that 
one gets if one includes isospin violation), the allowed 
7-tuples $(m,n,p,q,t,u,k)$  are listed along with the corresponding
terms. The main aim is to find the number of finite O($q^4)$
terms, given that the UV divergent terms have already been worked out in
\cite{mms}. For this purpose, the terms in tables 1 and 2
are labeled as F denoting the finite terms and D denoting the  
divergent terms. For the purpose of comparison with 
\cite{mms}, we have also indicated
which terms in table 1 of \cite{mms} the D-type terms correspond
to.  The LECs of O$(q^4)$ terms in
\cite{mms} are denoted by $d_i, i=1$ to 199. 
Further, the $i=188$ term in Table 1 of \cite{mms} should have
${\rm S}_\rho$ instead of $v_\rho$.

Overall, one gets 27 finite and 79 divergent (counter) terms
at O$(q^4)$.

\section{On-shell reduction}

In this section, we discuss the derivation of the on-shell O($q^4$)
${\cal L}_{\rm HBChPT}$, directly within HBChPT using
the techniques of \cite{1n}.

The main result obtained in \cite{1n} extended
to include external fields in the context of complete
on-shell reduction within HBChPT was the following rule:
\begin{eqnarray}
\label{eq:ruleonsh}
& & {\cal A}-{\rm type\  terms\ of\ the\ form\
{\bar{\rm H}}{\rm S}\cdot{\rm D}{\cal O}\rm H\ +\ h.c.} \nonumber\\
& & {\rm or\ {\bar{\rm H}}v\cdot{\rm D}{\cal O}\rm H\ +\ h.c.} \nonumber\\
& & {\rm or\ {\bar{\rm H}}{\cal O}^\mu\rm D_\mu\rm H\ +\ h.c.\ can\ be\
eliminated}\nonumber\\
& & {\rm except\ for}\
{\cal O}_\mu\equiv\Biggl(i^{m_1+l_5+l_7+t+u+1},\ {\rm or}\
\epsilon^{\nu\lambda\kappa\rho}\times\Omega\Biggr)\times u_\mu\Lambda\nonumber\\
& & {\rm with}\ l_1\geq1,
\Omega\equiv 1(i)\ {\rm for}\ (-)^{m_1+l_5+l_7+t+u+1}=-1(1),\nonumber\\
& & {\rm or}\nonumber\\
& & {\cal O}_\mu\equiv\Biggl(i^{m_1+l_5+l_7+k+t+u},\ {\rm or}\
\epsilon^{\nu\lambda\kappa\rho}\times\Omega^\prime\Biggr)
\times {\rm D}_\mu\Lambda,\nonumber\\
& & {\rm with}\ l_1\geq1, \Omega^\prime\equiv 1(i)\ {\rm for}\ 
(-)^{m_1+l_5+l_7+k+t+u}=-1(1).
\end{eqnarray}
In (\ref{eq:ruleonsh})
\begin{equation}
\label{eq:exception8}
\Lambda\equiv\prod_{i=1}^{M_1}{\cal V}_{\nu_i}
\prod_{j=1}^{M_2} u_{\rho_j}
(v\cdot u)^{l_1}u^{2l_2}\chi_+^{l_3}\chi_-^{l_4}
([v\cdot{\rm D},\ ])^{l_5}
({\rm D}_\beta{\rm D}^\beta)^{l_6}(u_\alpha{\rm D}^\alpha)^{l_7}
(v^{(s)}_{\sigma\omega})^k(F^+_{\rho\lambda})^t
(F^+_{\mu\nu})^u{\rm S}^r_\kappa,
\end{equation}
where ${\cal V}_{\nu_i}\equiv v_{\nu_i}\ \rm or\ \rm D_{\nu_i}$.
where ${\cal V}_{\nu_i}\equiv v_{\nu_i}\ \rm or\ \rm D_{\nu_i}$.
The number of ${\rm D}_{\nu_i}$s in (\ref{eq:exception8})
equals $m_1(\leq M_1)$.
Assuming that Lorentz invariance, parity and hermiticity
have been implemented, the choice of the factors
of $i$ in (\ref{eq:ruleonsh})
automatically incorporates the phase rule (\ref{eq:phoffsh}).
In (\ref{eq:ruleonsh}),
it is only the contractions of the building blocks that
has  been indicated. Also, traces have not been indicated.
It is  understood that all (anti-)commutators  in the HBChPT Lagrangian
are to be expanded out until one hits the first ${\rm D}_\mu$, so that the
${\cal A}$-type HBChPT term can be put in the form
${\bar{\rm H}}{\cal O}^\mu\rm D_\mu\rm H$
+h.c.

The complete on-shell O($q^4$) HBChPT Lagrangian can be shown
to be given by:
\begin{eqnarray}
\label{eq:onshO41}
& & {\cal L}_{\rm HBChPT}^{(4)} = A^{(4)}+
{1\over{2\rm m}}\gamma^0B^{(2)}\ ^\dagger\gamma^0 B^{(2)}\nonumber\\
& & +{1\over{2\rm m}}\biggl[\gamma^0 B^{(3)}\ ^\dagger\gamma^0 B^{(1)}
+\gamma^0B^{(1)}\ ^\dagger\gamma^0B^{(3)}\biggr]
\nonumber\\
& & -{1\over{4\rm m^2}}\biggl[\gamma^0 B^{(2)}\ ^\dagger\gamma^0 C^{(1)}B^{(1)}
+\gamma^0 B^{(1)}\ ^\dagger\gamma^0 C^{(1)}B^{(2)}\biggr]\nonumber\\
& & -{1\over{4\rm m^2}}\gamma^0 B^{(1)}\ ^\dagger\gamma^0 C^{(2)}B^{(1)}
+{1\over{8\rm m^3}}
\gamma^0 B^{(1)}\ ^\dagger\gamma^0(C^{(1)})^2B^{(1)}.
\end{eqnarray}

Using
\begin{equation}
\label{eq:B2onsh}
B^{(2)}_{\rm OS}[v^{(s)},F^+,\chi_-]=
\alpha_3\gamma^5\chi_3
i\alpha_7\gamma^5v^{[\mu}{\rm S}^{\nu]}v^{(s)}_{\mu\nu}
+i\alpha_8\gamma^5v^{[\mu}{\rm S}^{\nu]}F^+_{\mu\nu}
+\alpha_9\rangle\chi_-\rangle,
\end{equation}
(OS$\equiv$on-shell) one gets:
\begin{eqnarray}
\label{eq:B2B2onsh}
& & {1\over{2\rm m}}\gamma^0B^{(2)}\ ^\dagger\gamma^0 
B^{(2)}[v^{(s)},F^+,\chi_-]
\nonumber\\
& & =
{1\over{2\rm m}}\Biggl[
+(\alpha_3\chi_-+\alpha_9\langle\chi_-\rangle)^2
-\alpha_1[[v\cdot u,{\rm S}\cdot u],
(\alpha_-+\alpha_9\langle\chi_-\rangle)]_+
\nonumber\\
& & 
-v^{[\mu}{\rm S}^{\nu]}
[(\alpha_7v^{(s)}_{\mu\nu}+\alpha_8F^+_{\mu\nu},
(\alpha_3\chi_-+\alpha_9\langle\chi_-\rangle)]_+
+i\alpha_4[(\alpha_3\chi_-+\alpha_9\langle\chi_-\rangle),
[v\cdot{\rm D},v\cdot u]]_+
\nonumber\\
& &
+(\alpha_7v^{(s)}_{\mu\nu}+\alpha_8F^+_{\mu\nu}
(\alpha_7v^{(s) \nu}_\rho+\alpha_8F^{+\nu}_\rho)
-{i\over2}\epsilon^{\nu\lambda\alpha\beta}v_\alpha v^\mu
v^\rho
{\rm S}_\beta(\alpha_7v^{(s)}_{\mu\nu}+\alpha_8F^+_{\mu\nu})
(\alpha_7v^{(s)}_{\rho\lambda}+\alpha_8F^+_{\rho\lambda})
\nonumber\\
& & i\alpha_1\alpha_7
\biggl(-{1\over4}v^\mu[[v\cdot u,u^\nu],v^{(s)}_{\mu\nu}]_+\biggr)
+i\alpha_1\alpha_8
\biggl(-{1\over4}v^\mu[[v\cdot u,u^\nu],F^+_{\mu\nu}]_+ +{1\over2}
\epsilon^{\nu\rho\alpha\beta}v_\alpha{\rm S}_\beta v^\mu
[[v\cdot u,u_\rho],F^+_{\mu\nu}]\biggr)\nonumber\\
& & 
+i\alpha_2\alpha_7\biggl(
-{1\over2}\epsilon^{\nu\rho\alpha\beta}v_\alpha{\rm S}_\beta
v^\mu[[v\cdot u,u^\rho]_+,v^{(s)}_{\mu\nu}]_+\biggr)\nonumber\\
& & 
+i\alpha_2\alpha_8\biggl(
{1\over4}v^\mu[[v\cdot u,u^\nu]_+,F^+_{\mu\nu}]
-{1\over2}\epsilon^{\nu\rho\alpha\beta}v_\alpha{\rm S}_\beta
v^\mu[[v\cdot u,u^\rho]_+,F^+_{\mu\nu}]_+\biggr)\nonumber\\
& & -\alpha_3
v^{[\mu}{\rm S}^{\nu]}[(\alpha_7v^{(s)}_{\mu\nu}+\alpha_8F^+_{\mu\nu}),\chi_-]_+
-\alpha_4v^{[\mu}{\rm S}^{\nu]}[(\alpha_7v^{(s)}_{\mu\nu}+\alpha_8F^+_{\mu\nu}),
[v\cdot{\rm D},v\cdot u]]_+\nonumber\\
& &  -i\alpha_2\alpha_4[[v\cdot u,{\rm S}\cdot u]_+,[v\cdot{\rm D},
v\cdot u]]+i\alpha_3\alpha_4[\chi_-,[v\cdot{\rm D},v\cdot u]]_+\biggr].
\end{eqnarray}

Using (\ref{eq:B2onsh}) and $C^{(1)}_{\rm OS}$, 
and eliminating all terms proportional to the nonrelativistic
eom by  field redefinition of H, one gets:
\begin{eqnarray}
\label{eq:B2C1B1onsh}
& & -{1\over{4\rm m^2}}\gamma^0 B^{(2)}\ ^\dagger\gamma^0 C^{(1)}B^{(1)}
[v^{(s)},F^+]+{\rm h.c.}
\nonumber\\
& & 
=(\alpha_7v^{(s)}_{\mu\nu}
+\alpha_8F^+_{\mu\nu})\biggl[-2iv^\mu[v\cdot{\rm D},{\rm D}^\nu]
+g_A^0{\rm D}^\nu{\rm S}\cdot u+2\epsilon^{\nu\rho\lambda\sigma}
v^\mu v_\lambda{\rm S}_\sigma [v\cdot{\rm D},{\rm D}_\rho]\nonumber\\
& & -{i\over2}g_A^0\epsilon^{\nu\rho\lambda\sigma}v^\mu v_\lambda
{\rm D}_\rho u_\sigma-g_A^0v^\mu{\rm S}^\nu[v\cdot{\rm D},v\cdot u]
-{{g_A^0}^2\over4}v\cdot u u^\nu v^\mu\nonumber\\
& & +{i{g_A^0}^2\over2}
\epsilon^{\alpha\nu\rho\omega}v^\mu v_\rho{\rm S}_\omega v\cdot u u_\alpha
+g^0_A{\rm S}^\nu v^\mu{\rm D}\cdot u-g_A^0 v^\mu{\rm S}\cdot{\rm D} u^\nu
\nonumber\\
& & -g^0_Av^\mu{\rm S}^\nu[v\cdot{\rm D},v\cdot u]+{i\over2}
\epsilon^{\rho\lambda\alpha\nu}v^\mu v_\alpha u_\rho{\rm D}_\lambda
+g^0_Av^\mu u^\nu{\rm S}\cdot{\rm D}-g^0_A v^\mu {\rm S}\cdot u
{\rm D}^\nu\nonumber\\
& & 
+{i{g^0_A}^2\over2}(-{1\over2}v^\mu u^\nu+i\epsilon^{\nu\rho\lambda\sigma}
v^\mu v_\lambda{\rm S}_\sigma u_\rho)\biggr]\nonumber\\
& & +\biggl(2
(\alpha_3\chi_-+\alpha_9\langle\chi_-\rangle)v\cdot{\rm D}{\rm S}
\cdot{\rm D}
-i{g_A^0\over2}
(\alpha_3\chi_-+\alpha_9\langle\chi_-\rangle)[v\cdot{\rm D},v\cdot u]
+{{g_A^0}^2\over2}
(\alpha_3\chi_-+\alpha_9\langle\chi_-\rangle)v\cdot u
{\rm S}\cdot u\nonumber\\
& &
+g_A^0(\alpha_3\chi_-+\alpha_9\langle\chi_-\rangle)
\biggl[{1\over4}v\cdot u{\rm S}\cdot u+{i\over4}u_\mu{\rm D}^\mu
+{1\over2}
\epsilon^{\mu\nu\rho\lambda}v_\rho{\rm S}_\lambda u_\mu{\rm D}_\nu\biggr]
+{\rm h.c.}\Biggr]
\end{eqnarray}
Similarly, using:
\begin{equation}
\label{eq:A2onsh}
C^{(2)}_{\rm S}[v^{(s)},F^+,\chi_+]=-\alpha_6\chi_+
\alpha_7\epsilon^{\mu\nu\rho\lambda}v_\rho{\rm S}_\lambda
v^{(s)}_{\mu\nu}
+\alpha_8\epsilon^{\mu\nu\rho\lambda}v_\rho{\rm S}_\lambda
F^+_{\mu\nu} -\alpha_{10}\langle\chi_+\rangle
\end{equation}
and
\begin{equation}
\label{eq:B1}
B^{(1)}=-2i\gamma^5{\rm S}\cdot{\rm D}-{g_A^0\over2}\gamma^5v\cdot u,
\end{equation}
and eliminating all terms proportional to the nonrelativistic
eom by  field redefinition of H, one sees that:
\begin{eqnarray}
\label{eq:B1C2B1onsh}
& & -{1\over{4\rm m^2}}\gamma^0 B^{(1)}\ ^\dagger\gamma^0 C^{(2)}B^{(1)}
[v^{(s)},F^+,\chi_+]
\nonumber\\
& & =-{1\over{4\rm m^2}}\Biggl[
-\epsilon^{\mu\nu\rho\lambda}v_\rho{\rm S}\cdot{\rm D}
(\alpha_7v^{(s)}_{\mu\nu}+\alpha_8F^+_{\mu\nu})
{\rm D}_\lambda+\biggl(i{\rm D}^\mu
(\alpha_7v^{(s)}_{\mu\nu}+\alpha_8F^+_{\mu\nu}){\rm D}^\nu\nonumber\\
& & 
+g_A^0{\rm S}\cdot u
(\alpha_7v^{(s)}_{\mu\nu}+\alpha_8F^+_{\mu\nu})v^\mu{\rm D}^\nu
+g_A^0v^\mu{\rm D}^\nu
(\alpha_7v^{(s)}_{\mu\nu}+\alpha_8F^+_{\mu\nu})
{\rm S}\cdot u\biggr)\nonumber\\
& & -{{g_A^0}^2\over4}\epsilon^{\mu\nu\rho\lambda}v_\rho
v\cdot u
(\alpha_7v^{(s)}_{\mu\nu}+\alpha_8F^+_{\mu\nu})
{\rm S}_\lambda v\cdot u\nonumber\\
& & +2\epsilon^{\mu\nu\rho\lambda}v_\rho\biggl(-{ig_A^0\over4}v^\omega{\rm D}_\mu
(\alpha_7v^{(s)}_{\omega\nu}+\alpha_8F^+_{\omega\nu})u_\lambda
-{\rm D}_\mu
(\alpha_7v^{(s)}_{\omega\nu}+\alpha_8F^+_{\omega\nu}){\rm D}^\omega{\rm S}_\lambda
\biggr)\nonumber\\
& & 
-g^0_A\biggl({\rm S}\cdot{\rm D}
(\alpha_7v^{(s)}_{\mu\nu}+\alpha_8F^+_{\mu\nu})u^\nu
v^\sigma -{\rm D}^\rho v^\sigma
(\alpha_7v^{(s)}_{\mu\nu}+\alpha_8F^+_{\mu\nu})u_\rho\nonumber\\
& & 
+{ig_A^0\over4} u^\nu v^\sigma
(\alpha_7v^{(s)}_{\mu\nu}+\alpha_8F^+_{\mu\nu})v\cdot u
-{g_A^0\over2}\epsilon^{\mu\nu\rho\lambda}v_\rho{\rm S}_\lambda u_\mu
(\alpha_7v^{(s)}_{\mu\nu}+\alpha_8F^+_{\mu\nu})v^\sigma v\cdot u\nonumber\\
& & 
+\biggl({ig_A^0\over4}\epsilon^{\mu\nu\rho\lambda}v_\rho{\rm D}_\lambda
(\alpha_7v^{(s)}_{\mu\nu}+\alpha_8F^+_{\mu\nu})
v\cdot u-{i{g_A^0}^2\over4}v^\mu u^\nu
(\alpha_7v^{(s)}_{\mu\nu}+\alpha_8F^+_{\mu\nu})
v\cdot u\nonumber\\
& & 
-{{g_A^0}^2\over2}\epsilon^{\mu\nu\rho\lambda}v_\rho{\rm S}_\lambda v^\kappa u_\mu
(\alpha_7v^{(s)}_{\mu\nu}+\alpha_8F^+_{\mu\nu})v\cdot u\nonumber\\
& & 
-g_A^0{\rm D}^\nu
(\alpha_7v^{(s)}_{\mu\nu}+\alpha_8F^+_{\mu\nu}){\rm S}^\nu+h.c.\biggr)
\nonumber\\
& & 
+\Biggl({g_A^0}^2\biggl[{1\over 4}v\cdot u
(\alpha_6\chi_++\alpha_{10}\langle\chi_+\rangle)v\cdot u
-{1\over4}u_\mu
(\alpha_6\chi_++\alpha_{10}\langle\chi_+\rangle)u^\mu+{i\over2}
\epsilon^{\mu\nu\rho\lambda}v_\rho{\rm S}_\lambda u_\mu
(\alpha_6\chi_++\alpha_{10}\langle\chi_+\rangle)u_\nu\biggr]
\nonumber\\
& & +{\rm D}_\mu\chi_+{\rm D}^\mu
-2i\epsilon^{\mu\nu\rho\lambda}v_\rho{\rm S}_\lambda
{\rm D}_\mu\chi_+{\rm D}_\nu\nonumber\\
& & +(ig_A^0v\cdot u\chi_+{\rm S}\cdot{\rm D}+{\rm h.c.})
+{{g_A^0}^2\over 4}v\cdot u\chi_+v\cdot u\Biggr)
\Biggr]
\end{eqnarray}

Using: 
\begin{eqnarray}
\label{eq:B3onsh}
& & 
B^{(3)}_{\rm OS}[v^{(s)},F^+,\chi_+,\chi_-]=
\beta_4\gamma^5[v\cdot u,\chi_+]_+
+\beta_{14}\gamma^5[\chi_-,{\rm S}\cdot u]\nonumber\\
& & +\beta_{17}
\gamma^5 {\rm S}^\nu [{\rm D}^\mu,v^{(s)}_{\mu\nu}]
i\beta_{18}\gamma^5v^\mu[F^+_{\mu\nu},u^\nu]+\beta_{19}
\epsilon^{\mu\nu\rho\lambda}\gamma^5{\rm S}_\lambda
[F^+_{\mu\nu},u_\rho]_+\nonumber\\
& & 
+\beta_{20}\epsilon^{\mu\nu\rho\lambda}\gamma^5
v^{(s)}_{\mu\nu}u_\rho{\rm S}_\lambda
+\beta_{21}\epsilon^{\mu\nu\rho\lambda}\gamma^5v_\rho{\rm S}_\lambda
[F^+_{\mu\nu},v\cdot u]
\nonumber\\
& & 
+\beta_{22}\gamma^5{\rm S}^\nu[{\rm D}^\mu,F^+_{\mu\nu}]
+\beta_{23}\gamma^5\langle[v\cdot u,\chi_+]_+\rangle\nonumber\\
& & +\beta_{2}\gamma^5v\cdot u
\langle\chi_+\rangle+i\beta_{25}\gamma^5\langle[v\cdot{\rm D},
\chi_-]\rangle, 
\end{eqnarray}
${1\over{2\rm m}}\biggl[\gamma^0 B^{(3)}\ ^\dagger\gamma^0 B^{(1)}
+\gamma^0B^{(1)}\
^\dagger\gamma^0B^{(3)}\biggr][v^{(s)},F^+,\chi_-,\chi_+]$
and eliminating all terms proportional to the nonrelativistic
eom by  field redefinition of H, one gets:
\begin{eqnarray}
\label{eq:B3B1onsh}
& & {1\over{2\rm m}}\Biggl[
\beta_4\biggl(-2i[{\rm S}\cdot{\rm D},[v\cdot u,\chi_+]_+]_+
-{g_A^0\over2}[v\cdot u,[v\cdot u,\chi_+]_+]_+\biggr)\nonumber\\
& & +\beta_{14}\biggl(i[{\rm D}_\mu, [\chi_-,u^\mu]]_+
+g_A^0[{\rm S}\cdot u,[\chi_-,v\cdot u]]_+
-\epsilon^{\mu\nu\rho\lambda}v_\rho{\rm S}_\lambda
[{\rm D}_\nu,[\chi_-,u_\mu]]\nonumber\\
& &
-{g_A^0\over2}[v\cdot u,[\chi_-,{\rm S}\cdot u]]_+\biggr)
\nonumber\\
& & +\biggl({ig^0_A\over2}[{\rm D}^\mu,
(i\beta_{17}v^{(s)}_{\mu\nu}+i\beta_{22}F^+_{\mu\nu})]
v^\nu{\rm S}\cdot uu 
\nonumber\\
& & -{1\over2}[{\rm D}^\mu,
(i\beta_{17}v^{(s)}_{\mu\nu}+i\beta_{22}F^+_{\mu\nu})]{\rm D}_\nu\nonumber\\
& & +i\epsilon^{\nu\rho\lambda\sigma}v_\lambda{\rm S}_\sigma[{\rm D}^\mu,
(i\beta_{17}v^{(s)}_{\mu\nu}+i\beta_{22}F^+_{\mu\nu})]{\rm D}^\rho
\nonumber\\
& & 
-i{g_A^0\over2}{\rm S}^\nu[{\rm D}^\mu,
(i\beta_{17}v^{(s)}_{\mu\nu}+i\beta_{22}F^+_{\mu\nu})]v\cdot u\biggr)
\nonumber\\
& & +\beta_{18}\biggl(-2v^\mu[F^+_{\mu\nu},u^\nu]{\rm S}\cdot{\rm D}+{ig_A^0\over2}
v^\mu[F^+_{\mu\nu},u^\nu]v\cdot u\biggr)
+\beta_{19}\biggl({i\over2}\epsilon^{\lambda\alpha\rho\beta}[F^+_{\lambda\alpha},u_\rho]_+
(ig_A^0v_\beta {\rm S}\cdot u-{\rm D}_\beta)
\nonumber\\
& & -2([F^+_{\lambda\alpha},u_\nu]_+v^\lambda{\rm S}^\alpha{\rm D}^\nu
-{\rm S}^\alpha[F^+_{\lambda\alpha},v\cdot u]_+{\rm D}^\lambda
+v^\alpha[F^+_{\lambda\alpha},{\rm S}\cdot u]_+{\rm D}^\nu)
\nonumber\\
& & +{g^0_A\over2}\epsilon^{\lambda\alpha\rho\beta}{\rm S}_\beta
[F^+_{\lambda\alpha},u_\rho]_+v\cdot u\biggr)
+\beta_{20}\biggl({i\over2}\epsilon^{\lambda\alpha\rho\beta}
v^{(s)}_{\lambda\alpha} 
u_\rho (-ig_A^0{\rm S}\cdot u-{\rm D}_\beta)
+{g_A^0\over2}\epsilon^{\lambda\alpha\rho\beta}v^{(s)}_{\lambda\alpha}u_\rho 
v\cdot u \nonumber\\
& & -2(v^{(s)}_{\lambda\alpha}v^\lambda{\rm S}^\alpha u_\nu{\rm D}^\nu 
-{\rm S}^\lambda v^{(s)}_{\lambda\alpha}v\cdot u
{\rm D}^\lambda+v^\lambda v^{(s)}_{\lambda\alpha}
{\rm S}\cdot u{\rm D}^\nu)\biggr)\nonumber\\
& & +\beta_{21}\biggl({i\over2}\epsilon^{\lambda\alpha\rho\beta}v_\rho
[F^+_{\lambda\alpha},v\cdot u]{\rm D}_\beta
+2[F^+_{\lambda\alpha},v\cdot u]v^\lambda(-{1\over4}u^\alpha+{i\over2}
\epsilon^{\mu\nu\rho\kappa}v_\rho{\rm S}_\kappa u_\mu)\nonumber\\
& & 
-2{\rm S}^\alpha [F^+_{\lambda\alpha},v\cdot u]{\rm D}^\lambda
-{g^0_A\over2}\epsilon^{\lambda\alpha\rho\beta}v_\rho{\rm S}_\beta
[F^+_{\lambda\alpha},v\cdot u]v\cdot u\biggr)\nonumber\\
& & +\beta_{23}\biggl(2i\langle[v\cdot 
u,\chi_+]_+\rangle{\rm S}\cdot{\rm D}
+{g_A^0\over2}v\cdot u\langle[v\cdot
u,\chi_+]_+\rangle\nonumber\\
& & +\beta_{24}(v\cdot u\langle
\chi_+\rangle{\rm S}\cdot{\rm D}+{g_A^0\over2}(v\cdot u)^2
\langle\chi_+\rangle\biggr)\nonumber\\
& & +\beta_{25}\biggl(-2\langle
[v\cdot{\rm D},\chi_-{\rm S}\cdot{\rm D}
+i{g_A^0\over2}\langle[v\cdot{\rm D},\chi_-]
\rangle v\cdot u\Biggr)\Biggr].
\end{eqnarray}
The set $\{\beta_i\}$ can be related to the set $\{b_i\}$ of \cite{em}.

\section{Conclusion}
A complete list of O$(q^4)$ terms for off-shell nucleons was
obtained {\it working within HBChPT}
using a phase rule obtained in \cite {1n}, along
with reductions from algebraic identities.
We also obtain the on-shell O($q^4$) terms, again within the framework
of HBChPT.  
For off-shell nucleons, one gets a total of
106 ${\cal O}(q^4)$ terms (given in Tables 1 and 2).
Of these  27 are finite. 
Contrary to what is claimed in \cite{Fettes}, the earlier version
of this paper itself 
contained (overcomplete) list of terms {\it including} external
fields. Besides, the whole point of this paper (and that of 
\cite{O4ext0} and \cite{1n}) is to carry out the $1/\rm m$-reduction
without actually doing it; this is carried out by developing a method
of imposing charge conjugation invariance directly within the 
nonrelativistic framework in terms of a phase rule (\ref{eq:phoffsh})
that can be used directly within HBChPT. Also, once having obtained 
the list of nonrelativistic terms, 
constructing their relativistic counterparts (for reasons given in
\cite{Fettes}) can easily be done by following \cite{1n}.
We are getting fewer terms than the ones given in \cite{Fettes}. 
Also, instead of working with symmetrized and anti-symmetrised
commutators of $\rm D_\mu$ and $u_\nu$, we use (10) to eliminate
$F^-_{\mu\nu}$ altogether.

\begin{table} [htbp]
\centering
\caption{The Allowed Terms}
\begin{tabular} {|c|c|c|c|c|} \hline
$i$ & $(m,n,p,q,k,u.t)$ & Terms &  F($\equiv$Finite) & E\\
& & &  D($\equiv$Divergent)[$d_i$] & ON \\ \hline
1 & (4,0,0,0,0,0,0)   &
$(v\cdot\rm D)^4$ & D[$d_{197}$] & E  \\ \hline
2 & & ${\rm D}^4$ & F &   E \\ \hline
3 & (0,4,0,0,0,0,0) &   $(v\cdot u)^4$ & D[$d_5$] & ON  \\
\hline
4& & $u^4$ & D[$d_2$] & ON \\ \hline
5& & $[u_\mu,u_\nu]_+^2$ & D[$d_1$]  & ON \\ \hline
6& & $[u_\mu,v\cdot u]_+^2$ & D[$d_4$] & ON \\
\hline
7& (2,2,0,0,0,0,0) &  $v\cdot{\rm D} (v\cdot u)^2 v\cdot{\rm D}$ &
D[$d_{146}$] & E \\ \hline
8& & $[v\cdot{\rm D},(v\cdot u)]^2$ &  
D[$d_{128}$] &  ON \\ \hline
9& & $[v\cdot{\rm D},u_\mu]^2$ & D[$d_{133}$] & ON \\
\hline
10& & $[{\rm D}_\mu,(v\cdot u)]^2$ & D[$d_{135}$] & ON \\        
\hline
11& & $v\cdot{\rm D}[v\cdot u,u^\mu]_+\rm D_\mu+h.c.$ & D[$d_{148}]$] &
E \\ \hline
12 & & $[{\rm D}_\mu, u_\nu]^2$ & D[$d_{136}$]& ON \\ \hline
13& & ${\rm D}_\mu u^2{\rm D}^\mu$ & D[$d_{150}$] & E \\   
\hline
14& &
$[v\cdot u,[[v\cdot{\rm D},v\cdot u],v\cdot{\rm D}]]_+$ & D[$d_{129}$] &
ON
\\ \hline
15& & $[v\cdot{\rm D},[[v\cdot{\rm D},v\cdot u],v\cdot u]]_+$ &
D[$d_{138}$] & E \\ \hline
16& &  $[u_\mu,[[v\cdot{\rm D}, u^\mu], v\cdot{\rm D}]]_+$ & D[$d_{134}$]
& ON \\ \hline
17& & $[v\cdot{\rm D},[[v\cdot{\rm D},u_\mu], u^\mu]]_+$
& D[$d_{142}$] & E \\ \hline
18& & $[v\cdot u,[[{\rm D}^\mu,v\cdot u],{\rm D}_\mu]]_+$
& D[$d_{131}$] & ON \\ \hline
19& & $[u^\nu,[[{\rm D}_\mu, u_\nu]_+,{\rm D}^\mu]]$ & D[$d_{137}$]
& ON \\ \hline
20& & $[{\rm D}_\mu,[[{\rm D}^\mu, u_\nu], u^\nu]]_+$ & D[$d_{143}$]
& E \\ \hline
21& & $[{\rm D}_\mu,[[{\rm D}_\nu,u^\mu],u^\nu]]_+$ & D[$d_{145}$]
& E \\ \hline 
22& & $[u_\mu,[[{\rm D}^\mu,v\cdot u],v\cdot{\rm D}]]_+$ &
D[$d_{130}$] & ON \\ \hline
23& & $[v\cdot{\rm D},[[{\rm D}_\nu,v\cdot u],u^\nu]]_+$ &
D[$d_{144}$] & E \\ \hline
24& & $[[{\rm D}_\mu,v\cdot u],[v\cdot{\rm D},u^\mu]]_+$ &
D[$d_{132}$] & ON \\ \hline
25& & $i[{\rm D}_\mu,[\tilde\chi_-,u^\mu]]_+$
& D[$d_{123}$] & E \\
\hline
26 & & $i\epsilon^{\mu\nu\rho\lambda}
v_\rho{\rm S}_\lambda
[[{\rm D}_\kappa,u_\mu],[{\rm D}^\kappa,u_\nu]]$ & D[$d_{169}$] &
ON \\ \hline
27 & & $i\epsilon^{\mu\nu\rho\lambda}
v_\rho{\rm S}_\lambda
[{\rm D}_\kappa,[[{\rm D}^\kappa,u_\mu],u_\nu]_+]_+$ & D[$d_{175}$]
& E\\ \hline
28 & (2,0,1,0,0,0,0)  &
$[\rm D_\mu,[\rm D^\mu,\chi_+]]$ & F & E \\ \hline
29 & & $\rm
D_\mu\chi_+\rm D^\mu$ & F & E\\ \hline
30 & & $v\cdot\rm D\chi_+v\cdot\rm D$  & D[$d_{159,161}$] & E \\ \hline
31& & $[v\cdot\rm D,[v\cdot\rm D,\chi_+]]$ &  D[$d_{156,157}$] & E \\ \hline
32 & &
$i\epsilon^{\mu\nu\rho\lambda}
v_\rho\rm S_\lambda[\rm D_\mu,[\rm D_\nu,\chi_+]_+]$ & F & E \\ \hline   
33 & (0,2,1,0,0,0,0)  & $u_\mu\chi_+u^\mu$ & D[$d_{10,11,12}$] & ON \\ \hline
34 & & $u^2\chi_+$ &  D[$d_{10,11}$] & ON \\ \hline
35& & $v\cdot u\chi_+v\cdot u$ & D[$d_{13,14,15}$] & ON \\ \hline
36
& & $(v\cdot u)^2\chi_+$ & D[$d_{13,14}$] & ON   \\ \hline
\end{tabular}
\end{table}

\addtocounter{table}{-1}
\begin{table} [htbp]
\centering
\caption{continued}
\begin{tabular} {|c|c|c|c|c|} \hline
$i$ & $(m,n,p,q)$ & Terms &  F($\equiv$Finite) & E\\
& & &  D($\equiv$Divergent)[$d_i$] & ON \\ \hline 
37& &
$i\epsilon^{\mu\nu\rho\lambda}v_\rho{\rm S}_\lambda[[u_\mu,u_\nu],\chi_+]_+$
& F & ON \\ \hline
38 & (0,0,2,0,0,0,0) & $\chi_+^2$       &  D[$d_{21}$] & ON \\ \hline
39 &
(0,0,0,2,0,0,0) & $\chi_-^2$ & F & ON \\ \hline
40& (0,0,1,0,1,0,0) &
$\epsilon^{\mu\nu\rho\lambda}v_\rho{\rm S}_\lambda
F^+_{\mu\nu}\chi_+$ & D[$d_{54,56}$]& ON\\ \hline
41& (0,0,0,0,2,0,0) & $(F^+_{\mu\nu})^2$ & F & ON \\ \hline 
42& & $v^\kappa v^\sigma [F^+_{\mu\kappa},F^{+\mu}_\sigma]_+$ &
D[$d_{18}$] & ON \\ \hline
43 & & $\epsilon^{\mu\nu\rho\lambda}
v_\rho{\rm S}_\lambda
[F^+_{\mu\kappa},F^{+\kappa}_\nu]$ & D[$d_{52}$] & ON \\ \hline
 44& (2,0,0,1,0,0) &
$i[{\rm D}^\mu,[F^+_{\mu\nu},{\rm D}^\nu]]_+$ & D[$d_{155}$] &
E \\ \hline
45& & $\epsilon^{\mu\nu\rho\lambda}
v_\rho{\rm S}_\lambda
[[F^+_{\mu\kappa},{\rm D}^\kappa],{\rm D}_\nu]$
& D[$d_{187}$] & E \\ \hline
46& & $iv^\mu[[F^+_{\mu\nu},v\cdot{\rm D}],{\rm D}^\nu]_+$ & D[$d_{153}$]
& E
\\ \hline 
47& & $\epsilon^{\mu\nu\rho\lambda}
v_\rho{\rm S}_\lambda
[{\rm D}_\kappa,[F^+_{\mu\nu},{\rm D}^\kappa]]$ & D[$d_{181}$] & E \\
\hline
48& & $\epsilon^{\mu\nu\rho\lambda}
v_\rho{\rm S}_\lambda {\rm D}_\kappa F^+_{\mu\nu}{\rm D}^\kappa$ &
D[$d_{184}$] & E \\ \hline
49
& (0,0,1,0,0,0,1)
& $\epsilon^{\mu\nu\rho\lambda}v_\rho{\rm S}_\lambda
v^{(s)}_{\mu\nu}\chi_+$ & D[$d_{55}$] & ON \\ \hline
50 & (0,0,0,0,1,0,1)
& $v^{(s)}_{\mu\nu}F^{+\mu\nu}$ & D[$d_{17}$] & ON \\ \hline
51 & &
$v^\kappa v^\sigma v^{(s)}_{\kappa\mu}F^{+\ \mu}_\sigma$ & D[$d_{19}$] &
ON \\ \hline
52& (0,2,0,0,1,0,0) &
$i[u_\mu,[ F^+_{\mu\nu},u_\nu]]_+$ & F & ON \\ \hline
53& & $\epsilon^{\mu\nu\rho\lambda}v_\rho{\rm S}_\lambda
[F^+_{\mu\kappa},[u^\kappa,u_\nu]_+]_+$ & D[$d_{42}$] & ON \\ \hline
54& & $\epsilon^{\mu\nu\rho\lambda}v_\rho{\rm S}_\lambda
[[F^+_{\mu\kappa},u_\nu]_+,u^\kappa]_+$ & D[$d_{41}$] & ON\\ \hline
55& & $\epsilon^{\mu\nu\rho\lambda}v_\rho{\rm S}_\lambda
[[F^+_{\mu\kappa},u^\kappa]_+,u_\nu]_+$ & D[$d_{40}$] & ON \\ \hline
56& & $\epsilon^{\mu\nu\rho\lambda}
v_\rho{\rm S}_\lambda u_\kappa F^+_{\mu\nu}u^\kappa$ & D[$d_{39}$] & ON
\\ \hline
57& & $\epsilon^{\mu\nu\rho\lambda}v_\rho{\rm S}_\lambda F^+_{\mu\nu}
u^2$ & D[$d_{37}$] & ON \\ \hline
58& & $iv^\mu [[{\rm D}^\nu, v^{(s)}_{\mu\nu}],v\cdot{\rm D}]_+$ &
D[$d_{151}$ & E \\ \hline    
59& & $\epsilon^{\mu\nu\rho\lambda}v_\rho{\rm S}_\lambda
v^{(s)}_{\mu\nu}u^2$ & D$[d_{38}$] & ON \\ \hline
60& (3,1,0,0,0,0,0) &  
$i[v\cdot{\rm D},[v\cdot{\rm D},[v\cdot{\rm D},{\rm S}\cdot u]]]_+$
& D[$d_{190}$] & E \\ \hline
61 & & $i[v\cdot u,{\rm S}\cdot u]_+v\cdot u v\cdot{\rm D}+$ h.c.
& D[$d_{91}$] & E \\ \hline
 62 & (1,1,1,0,0,0,0) &
$i[[v\cdot{\rm D},{\rm S}\cdot u]_+,\chi_+]_+$ & D[$d_{117}$] & E \\
\hline
63& & $i[[{\rm S}\cdot{\rm D},v\cdot u]_+,\chi$
&  D[$d_{118}$] & E \\ \hline
64 &  (1,1,0,0,1,0,1)
& $v^{[\mu}{\rm S}^{\nu]}[F^+_{\mu\nu},v\cdot{\rm D}],v\cdot u]_+$
& D[$d_{99}$] & ON \\
\hline
65& & $v^{[\mu}{\rm S}^{\nu]}[F^+_{\mu\nu},[v\cdot{\rm D},v\cdot u]]_+$
& D[$d_{101}$] & ON \\ \hline
66& & $v^{[\mu}{\rm S}^{\nu]}[v\cdot{\rm D},[F^+_{\mu\nu},v\cdot u]]_+$
& D[$d_{108}$] & E
\\ \hline
67& & $v^{[\mu}{\rm S}^{\nu]}[[v^{(s)}_{\mu\nu},v\cdot{\rm D}],v\cdot
u]_+$
& D[$d_{94}$] & ON \\ \hline
68& & 
$v^{[\mu}{\rm S}^{\nu]}[v^{(s)}_{\mu\nu},[v\cdot{\rm D},v\cdot u]]_+$   
& D[$d_{98}$] & ON \\ \hline
\end{tabular}
\end{table}

\begin{table} [htbp]
\centering
\caption{The terms that need to be included if assume isospin violation}   
\begin{tabular} {|c|c|c|c|c|} \hline
$i$ & $(m,n,p,q,k,u.t)$ & Terms &  F($\equiv$Finite) & E\\
& & &  D($\equiv$Divergent)[$d_i$] & ON \\ \hline
69 & (2,0,0,,1,0,0,0) & $\langle[{\rm D}_\mu,[{\rm D}^\mu,\chi_+]]
\rangle$ & D[$d_{158}]$ & ON \\ \hline
70 & & ${\rm D}_\mu\langle\chi_+\rangle\rm D^\mu$ & D[$d_{160}$] & E \\
\hline
71 &  & $\langle[v\cdot{\rm D},[v\cdot{\rm D},\chi_+]]
\rangle$ & D[$d_{157}]$ & ON \\ \hline
72 & & $v\cdot{\rm D}\langle\chi_+\rangle v\cdot\rm D$
& D$[d_{159}]$ & E \\ \hline
73 & (0,2,0,1,0,0,0) & $u^2\langle\chi_+\rangle$ & D[$d_{10}$]
& ON \\ \hline
74 & & $u^\mu\langle[u_\mu,\chi_+]_+\rangle$ & D[$d_{12}$] & ON
\\ \hline
75 & & $(v\cdot u)^2\langle\chi_+\rangle$ & D[$d_{13}$] & ON \\ \hline
76 & & $v\cdot u\langle[v\cdot u,\chi_+]\rangle$ & D[$d_{15}]$ & ON
\\ \hline
77 & & $i\epsilon^{\mu\nu\rho\lambda}v_\rho{\rm S}_\lambda
\langle[[u_\mu,u_\nu],\chi_+]_+\rangle$ & D[$d_{50}$ & ON \\ \hline
78 & & $i\epsilon^{\mu\nu\rho\lambda}v_\rho{\rm S}_\lambda
[u_\mu,u_\nu] \langle \chi_+ \rangle$ & D[$d_{51}]$ & ON \\ \hline
79 & & $i
\epsilon^{\mu\nu\rho\lambda}v_\rho {\rm S}_\lambda
[u_\mu,[u_\nu,\chi]_+]_+$ & F & ON \\ \hline
80& (0,0,1,0,1,0,0) &
$\epsilon^{\mu\nu\rho\lambda}v_\rho {\rm S}_\lambda
F^+_{\mu\nu}\langle\chi_+\rangle$ & D[$d_{54,55}]$
& ON \\ \hline
81& (0,0,1,0,0,1,0) &
$\epsilon^{\mu\nu\rho\lambda}v_\rho {\rm S}_\lambda
v^{(s)}_{\mu\nu}\langle\chi_+\rangle$ & D$[d_{55}]$  
& ON \\ \hline
82& (0,0,0,2,0,0,0) & $\chi_+\langle\chi_+\rangle$ &
D[$d_{20,21}$] & ON \\ \hline
83& & $\langle\chi_+\rangle^2$ & D[$d_{21}$] & ON \\
\hline   
84& (0,0,0,2,0,0,0) & $\chi_-\langle\chi_-\rangle$ &
F & ON \\ \hline
85& & $\langle\chi_-\rangle^2$ & F & ON \\
\hline
86& (1,1,0,1,0,0,0) & $i[\chi_+,[v\cdot{\rm D},{\rm S}\cdot u]]$
& D[$d_{115}$] & ON \\ \hline
87& & $i[[\chi_+,v\cdot{\rm D}],{\rm S}\cdot u]$
& D[$d_{116}$] & ON \\ \hline
88&  & $i[\chi_+,[{\rm S}\cdot{\rm D},v\cdot u]]$   
& F & ON \\ \hline
89 & & $i[[\chi_+,{\rm S}\cdot{\rm D}],v\cdot u]$ &   
F & ON \\ \hline
90& &
$i[v\cdot{\rm D},\langle[{\rm S}\cdot u,\chi_+]_+\rangle]_+$
& D[$d_{119}$] & E \\ \hline
91& &
$i[{\rm S}\cdot{\rm D},\langle[v\cdot u,\chi_+]_+\rangle]_+$
& F & E \\ \hline
92& & $i\langle\chi_+\rangle[{\rm S}\cdot{\rm D},v\cdot u]_+$ &
D[$d_{119}$] & ON \\ \hline
93& & $i[\langle\chi_+\rangle[v\cdot{\rm D},{\rm S}\cdot u]_+$ &
D[$d_{117}$] & ON \\ \hline
94& 
(1,1,0,0,1,0,0)  & $i\langle\chi_-\rangle [{\rm D}_\mu,u^\mu]$ & F & ON
\\ \hline
95& & $i [{\rm D}_\mu,[u^\mu, \langle\chi_-\rangle]_+]$
& F & E \\ \hline
96& & $i\langle\chi_-\rangle [v\cdot {\rm D},v\cdot u]$ & F & ON
\\ \hline
97& & $i [v\cdot{\rm D},[v\cdot u, \langle\chi_-\rangle]_+]$
& F & E \\ \hline
98& & $\epsilon^{\mu\nu\rho\lambda}v_\rho{\rm S}_\lambda
[{\rm D}_\mu,\langle[u_\nu,\chi_-]\rangle]_+$
& F & E \\ \hline
99& (0,2,0,0,1,0,0,0) &
$[[v\cdot u,{\rm S}\cdot u],\chi_-]_+$ & F & ON \\ \hline
100& & $[v\cdot u,[{\rm S}\cdot u,\chi_-]]_+$ & F & ON \\ \hline
101
& & $[{\rm S}\cdot u,[v\cdot u,\chi_-]]_+$ & F & ON \\ \hline
102& & $[[v\cdot u,{\rm S}\cdot u]\langle\chi_-\rangle$
& F & ON \\ \hline
103& (2,0,0,1,0,0,0) & $i[v\cdot{\rm D},[{\rm S}\cdot{\rm D},
\langle\chi_-\rangle]_+]$ & F & E \\ \hline
104& (2,0,0,1,0,0,0) & 
$i[v\cdot{\rm D},[{\rm S}\cdot{\rm D},\langle\chi_-\rangle]_+]$
& F & E \\ \hline
105& (0,0,0,1,1,0,0) & $iv^{[\mu}{\rm S}^{\nu]}F^+_{\mu\nu}  
\langle\chi_-\rangle$ & F & ON \\ \hline
106& (0,0,1,,0,0,1) & $iv^{[\mu}{\rm S}^{\nu]}
v^{(s)}_{\mu\nu}\langle\chi_-\rangle$ & F & ON \\ \hline
\end{tabular}
\end{table}

\end{document}